\documentclass[11pt]{article}
\usepackage{fullpage, times}
\newcommand{\defeq}{\stackrel{\Delta}{=}}
\newcommand{\PAR}{\mbox{{\sf PAR}}}
\newcommand{\GT}{\mbox{{\sf GT}}}
\newcommand{\Tr}{\mbox{{\rm Tr} }}

\newcommand{\ellone}[1]{\| #1 \|_1}
\newcommand{\E}{\mathop{\rm {\bf E}}}

\newcommand{\cH}{{\cal H}}
\newcommand{\cK}{{\cal K}}
\newcommand{\cI}{{\cal I}}
\newcommand{\cM}{{\cal M}}
\newcommand{\cP}{{\cal P}}

\newcommand{\cQ}{{\cal Q}}
\newcommand{\cX}{{\cal X}}
\newcommand{\cY}{{\cal Y}}
\newcommand{\cZ}{{\cal Z}}
\newcommand{\ket}[1]{| #1 \rangle}
\newcommand{\ketbra}[1]{| #1 \rangle \langle #1 |}
\newcommand{\braket}[2]{\langle #1 | #2 \rangle}
\newcommand{\bzero}{{\mathbf 0}}

\newtheorem{definition}{Definition} % Specify Definition
\newtheorem{theorem}{Theorem}    % Specify Theorem
\newtheorem{result}{Result}    % Specify Result
\newtheorem{fact}{Fact}    % Specify Fact
\newtheorem{lemma}{Lemma}      % Specify Lemma
\newenvironment{proof}{\noindent{\bf Proof:}}{\qed} % Specify proof
\newcommand{\qed}{\hfill{$\rule{6pt}{6pt}$}} %Box at end of proof

\title{Lower bounds for predecessor searching
in the cell probe model\footnote{
This paper is the journal version of the conference papers
\cite{sen:quantcell} and \cite{sen:pred}.
}
}

\author{
Pranab Sen\thanks{
Department of Combinatorics and Optimization, 
University of Waterloo, 
Waterloo, Ontario N2L 3G1, Canada. 
Email:  {\sf p2sen@iqc.ca}.
Work done while the author was a graduate student at TIFR, Mumbai,
India visiting UC Berkeley and DIMACS under a Sarojini Damodaran
International Fellowship grant, and a
postdoctoral researcher at LRI, Orsay, France supported by 
the EU 5th framework program QAIP IST-1999-11234 and
by CNRS/STIC 01N80/0502 and 01N80/0607 grants.
}
\and
S~.Venkatesh\thanks{
Algorithms and Complexity Group, Max-Planck Institut f\"{u}r
Informatik, Stuhlsatzenhausweg 85, 66123 Saarbr\"{u}ken, Germany.
Email: {\sf srini@mpi-sb.mpg.de}. Work done while the author
was a postdoctoral researcher at IAS, Princeton, USA supported
by NSF grant CCR-9987845 and a joint IAS-DIMACS post doctoral
fellowship.
}
}

\date{}

\begin{document}

\maketitle
\thispagestyle{empty}

\begin{abstract}
We consider a fundamental problem in data structures, 
{\em static predecessor searching}: Given a subset $S$ of 
size $n$ from the universe 
$[m]$, store $S$ so that queries of the form ``What 
is the predecessor of $x$ in
$S$?'' can be answered efficiently. We study this problem in 
the cell probe 
model introduced by Yao~\cite{yao:tablesort}. Recently,
Beame and Fich~\cite{beame:pred} obtained optimal bounds on the 
number of probes
needed by any deterministic query scheme if the associated storage
scheme uses only $n^{O(1)}$ cells of word size $(\log m)^{O(1)}$ bits.

We give a new lower bound proof for this problem that matches 
the bounds of Beame and Fich. Our lower bound proof has 
the following advantages: it 
works for randomised query schemes too, while 
Beame and Fich's proof works for deterministic query schemes only. 
In addition, it is simpler than 
Beame and Fich's proof.
In fact, our lower bound for predecessor searching
extends to the `quantum address-only' query schemes that we
define in this paper. In these query schemes, quantum parallelism 
is allowed 
only over the `address lines' of the queries. These query schemes
subsume classical randomised query schemes,
and include many quantum query algorithms
like Grover's algorithm~\cite{grover:search}.

We prove our lower bound using the round elimination approach of 
Miltersen, Nisan, Safra and Wigderson~\cite{miltersen:roundelim}. 
Using tools from information theory, we prove a strong
round elimination lemma for communication complexity that enables 
us to
obtain a tight lower bound for the predecessor problem. 
Our strong round elimination lemma also extends to quantum
communication complexity. 
We also use our
round elimination lemma to obtain a rounds versus communication 
tradeoff for the `greater-than' problem, improving on the 
tradeoff in \cite{miltersen:roundelim}. 
We believe that
our round elimination lemma is of independent interest and should have
other applications.
\end{abstract}

%\noindent {\bf Keywords:} Predecessor searching, cell probe
%model, data structures, communication complexity, 
%information theory, lower bounds.

\section{Introduction}
\subsection{The problem and the model}
A static data structure problem
consists of a set of data $D$, a set of queries $Q$, a set of
answers $A$, and a function
$f:D \times Q \rightarrow A$. The aim is to store the data efficiently
and succinctly,
so that any query can be answered with only a few probes to the data
structure.
{\em Static predecessor searching}
is a well studied problem in data 
structure design (see e.g.~\cite{boas:priority, boas:order, 
willard:loglog, andersson:lin, ajtai:pred, miltersen:union, 
miltersen:roundelim, beame:pred}). 
Data structures for answering predecessor queries can be used to 
construct data structures to
answer other queries like rank (finding the number of elements in $S$ 
that are smaller than or equal to $x$) and nearest neighbour 
(finding an 
element in $S$ closest to $x$) efficiently. This motivates the need to
design efficient data structures that support predecessor queries. 

Let $[m]$ denote the set of integers $\{0, \ldots, m-1\}$.
\begin{definition}[Static predecessor searching]
In the problem of {\em $(m,n)$-static predecessor searching},
we are given a subset $S$ of size $n$ from the universe 
$[m]$. Our goal is to store the set $S$ succinctly
so that
queries of the form ``What is the predecessor of $x$ in $S$?'' for 
$x \in [m]$ can be answered with a few probes to the data structure. 
If $x$ has no predecessor
in $S$, that is, $x$ is smaller than 
every element in $S$, then return a default value, say, $-1$. 
\end{definition}

In this paper, we study the static predecessor searching problem in
Yao's {\em cell probe} model~\cite{yao:tablesort}.
The cell 
probe model is a natural and general model for proving upper and lower
bound results in data structures.
\begin{definition}[The cell probe model]
An $(s,w,t)$ {\em cell probe scheme} for a 
static data structure problem $f:D \times Q \rightarrow A$
has two components:
a {\em storage scheme} and a {\em query scheme}. The storage scheme
stores the data $d \in D$ as a table $T[d]$ of $s$
cells, each cell of word size $w$ bits.
The storage scheme is classical deterministic.
Given a query $q \in Q$, the query scheme computes $f(d, q)$ by
making at most $t$ probes to $T[d]$, where each probe reads one cell 
at a time, and the probes can be adaptive.
In a deterministic cell probe scheme the query scheme is 
classical deterministic,
in a randomised cell probe scheme it is classical randomised, and
in an address-only quantum cell probe scheme it is bounded error
address-only quantum (defined in Section~\ref{subsec:quantcell}).
\end{definition}
Since in the cell probe model we only charge a
scheme for the number of probes made to memory cells and for the
total number of cells of storage used, and all internal computation
is for free, 
lower bounds proved in the cell probe model hold in all reasonable
data structure models (e.g. the unit cost RAM with the same word
size) and give us insight into the intrinsic difficulty
of the problem.

The goal is to design cell probe schemes for $(m, n)$-static
predecessor searching using
small space i.e. $s = n^{O(1)}$ and $w = O(\log m)$, and at the
same time making a small number of probes $t$ in the worst case.

\subsection{Previous work}
We start by describing the sequence of results that lead to the 
currently
best known upper bounds for the $(m, n)$-static predecessor problem. 
For a long time, the best upper bound known for the 
predecessor problem
was due to the data structures of van Emde 
Boas et al.~\cite{boas:priority, boas:order}, 
and the data structures of
Fredman and Willard~\cite{fredman:fusion}. 
In their papers, van Emde 
Boas et al.~\cite{boas:priority, boas:order} 
gave a $(\Omega(m), O(\log m), O(\log \log m))$ deterministic
cell probe solution for predecessor. 
The main drawback of their solution is that the number of cells
used is very large. Later, Willard~\cite{willard:loglog} 
reduced the number of cells used to $O(n)$. 
Building on the work of van Emde Boas et al. and Willard, Fredman and
Willard~\cite{fredman:fusion}, and
Andersson~\cite{andersson:lin} designed 
$(O(n), O(\log m), O(\sqrt{\log n}))$
deterministic cell probe schemes for predecessor.
Recently, Beame and 
Fich~\cite{beame:pred} improved on these upper
bounds and showed a 
$\left(O\left(\frac{n^2 \log n}{\log \log n}\right),
       O(\log m),t\right)$,
where
$t = \min\left\{O\left(\frac{\log\log m}{\log\log\log m}\right),
O\left(\sqrt{\frac{\log n}{\log \log n}}\,\right)\right\}$,
deterministic cell probe scheme for predecessor. 

The first lower bound for the $(m, n)$-static predecessor
problem was proved by Ajtai~\cite{ajtai:pred}, who showed that
no $(n^{O(1)}, O(\log m), t)$ deterministic cell probe scheme for
predecessor can have constant number of probes $t$.
Miltersen~\cite{miltersen:union} observed that there 
is a close connection  
between the cell probe complexity of a data structure problem and
the communication complexity of a related communication game, and 
used this to improve Ajtai's lower bound to 
$\Omega(\sqrt{\log\log m}\,)$ probes. 
Recently, building on Ajtai's and Miltersen's work, 
Beame and Fich~\cite{beame:pred} 
showed that their data structure described above is indeed optimal
in the following sense: any 
$(n^{O(1)}, 2^{(\log m)^{1 - \Omega(1)}}, t)$ deterministic cell
probe scheme for predecessor must satisfy
$t = \Omega\left(\frac{\log\log m}{\log\log\log m}\right)$ 
as a function of $m$, and any 
$(n^{O(1)}, (\log m)^{O(1)}, t)$ deterministic cell
probe scheme for predecessor must satisfy
$t = \Omega\left(\sqrt{\frac{\log n}{\log \log n}}\,\right)$
as a function of $n$. Similar lower bounds
were proved by Xiao~\cite{xiao:cell}. We would like to
stress here that all the above lower bound proofs
are complicated with many
technical details.  Also, they hold for deterministic cell
probe schemes only.

The result of Beame and Fich gives rise to the following two 
questions: does their lower bound hold for randomised query
schemes as well? It has been observed recently that randomisation
enormously helps in the case of membership 
queries~\cite{buhrman:bitvectors} and
approximate nearest neighbour 
queries~\cite{liu:ann, kushilevitz:ann, indyk:ann, chakrabarti:ann}, 
and one might believe that it
could help answer predecessor queries quickly as well. Secondly, 
is it possible to give a simple proof of the lower bound result
of Beame and Fich?

A partial answer to both of the above questions was given by 
Miltersen, Nisan, Safra and Wigderson~\cite{miltersen:roundelim}. 
In their paper, they
proved a general {\em round elimination} lemma for communication 
complexity.
Using the connection between cell probe
complexity of data structures and communication complexity, and
their round elimination lemma, they showed 
the following lower bound for the 
predecessor problem: any $(n^{O(1)}, (\log m)^{O(1)}, t)$ randomised
cell probe scheme for $(m, n)$-static predecessor must
satisfy $t = \Omega(\sqrt{\log\log m}\,)$ 
as a function of $m$, and $t = \Omega((\log n)^{1/3})$
as a function of $n$. Though the 
lower bound proved by \cite{miltersen:roundelim} is 
weaker than that of 
\cite{beame:pred}, their approach had two advantages: their lower
bound holds for randomised query schemes too,
and the proof is much simpler.
In their paper, Miltersen et al. ask if their round elimination
based approach can be strengthened
to obtain the lower bound of Beame and Fich.

\subsection{Our Results}
We answer the question posed by Miltersen, Nisan, Safra and Wigderson.
Our main result in this paper shows that the lower bound of Beame and
Fich holds for address-only quantum cell probe schemes (and hence,
for randomised cell probe schemes) as well.

\subsubsection{The Predecessor Problem}
\begin{result}
Suppose there is a $(n^{O(1)}, (\log m)^{O(1)}, t)$ randomised 
cell probe
scheme for the $(m, n)$-static predecessor problem with error 
probability less than $1/3$.
Then, 
\begin{enumerate}
\item[(a)]
$t = \Omega \left( \frac{\log \log m}{\log \log \log m} \right)$ 
as a function of $m$; 
\item[(b)]
$t = \Omega \left( \sqrt{\frac{\log n}{\log \log n}} \, \right)$ 
as a function of $n$.
\end{enumerate}
The same lower bound also holds for address-only quantum cell 
probe schemes for static predecessor searching.
\end{result}

We prove our lower bound for predecessor searching 
by combining the approach in \cite{miltersen:roundelim}
with a new round elimination lemma for communication complexity.
Our round elimination lemma is a strengthening of the one proved 
in ~\cite{miltersen:roundelim},
and we believe it is of independent interest.

\subsubsection{An improved round elimination lemma}
In this paper, all communication protocols are two-party. 
The error probability of a randomised or quantum communication
protocol is defined as the maximum error of the 
protocol for any input.
For a general introduction to (classical) 
communication complexity, see the
book by Kushilevitz and Nisan~\cite{kushilevitz:cc}.

Let $f: \cX \times \cY \rightarrow \cZ$ be any 
communication problem. Let
us denote by $f^{(n),A}$ a new communication game in which
Alice is given $x_1,x_2,\ldots,x_n \in \cX$, and Bob
is given $y \in \cY$, $i \in [n]$ and also  
copies of $x_1, x_2, \ldots, x_{i-1}$. 
Their task is to compute $f(x_i,y)$. $f^{(n), B}$ is defined
similarly. Intuitively,
if Alice starts the communication and her first message
is much smaller than $n$ bits, then she is unlikely to send 
much useful information about $x_i$
to Bob as she is unaware of $i$. So it should be possible to 
eliminate the first message of Alice, giving rise to a protocol
where Bob starts, with one less round of communication, and having
similar message complexity and error probability.
The round elimination lemma captures
this intuition. 

\begin{definition}
A $[t;l_1,l_2,\ldots,l_t]^A$ ($[t;l_1,l_2,\ldots,l_t]^B$)
communication protocol is one
where Alice (Bob) starts the communication, the 
$i$th message is $l_i$ bits long, and the communication goes on
for $t$ rounds. 
\end{definition}

\begin{result}
Suppose the communication game $f^{(n),A}$ has a 
$[t;l_1,l_2,\ldots,l_t]^A$
public coin randomised protocol with error less than 
$\delta$. Then, $f$ has a $[t-1;l_2,\ldots,l_t]^B$ public coin 
randomised protocol 
with error less than 
$\epsilon \defeq \delta + (1/2)(2l_1\ln 2/n)^{1/2}$.
A similar result holds for public coin quantum protocols (defined
in Section~\ref{subsec:quantcomm}) too.
\end{result}

The proof of this lemma uses tools from information theory. 
In particular,
we use the {\em average encoding} theorem of 
Klauck, Nayak, Ta-Shma and Zuckerman~\cite{klauck:interaction}. 
Intuitively, this theorem 
says that if the mutual information between a random variable and its 
randomised encoding is small, then the probability distributions 
on code words for various values of the random variable
are indeed close to the average probability distribution on 
code words. 

\subsubsection{Applications to other problems}
We prove our lower bound result for predecessor by actually proving
a lower bound for the {\em rank parity problem}. In the rank parity
problem, we need to store a subset $S$ of the universe $[m]$ so that
given a query element $x \in [m]$, we can output whether the number
of elements in $S$ less than or equal to $x$ is even or odd.
Lower bounds for rank parity imply similar lower bounds for some other
data structure problems like {\em point separation}~\cite{beame:pred}
and {\em two-dimensional reporting 
range query}~\cite{miltersen:roundelim}. For
details of the reduction from rank parity to the above problems,
see the respective papers cited above.

Independently, the round elimination lemma has applications 
to problems 
in communication complexity. For example, let us consider 
communication protocols for the `greater-than' problem $\GT_n$ 
in which Alice and Bob are given bit strings $x$ and $y$ respectively
of length $n$ each, and 
the goal is to find out if $x > y$ or not (treating
$x, y$ as integers between $0$ and $2^n - 1$). 
Miltersen, Nisan, Safra and Wigderson~\cite{miltersen:roundelim},
and Smirnov~\cite{smirnov:shannon} have
studied rounds versus communication tradeoffs for $\GT_n$.
Miltersen et al. show an
$\Omega\left(n^{1/t} 2^{-O(t)}\right)$ lower bound for $t$-round
bounded error public coin randomised protocols for $\GT_n$.
Using our stronger round elimination lemma, we improve Miltersen
et al.'s result.
\begin{result}
The bounded error public coin randomised $t$-round
communication complexity of $\GT_n$ 
is lower bounded by $\Omega(n^{1/t}t^{-2})$.
For bounded error quantum protocols with input-independent
prior entanglement for
$\GT_n$, we have a lower bound of $\Omega(n^{1/t}t^{-1})$.
\end{result}

\paragraph{Remark:} The lower
bound for quantum protocols is better because, by definition, a
quantum protocol always sends messages whose length is independent 
of the input.

There exists a bounded error classical randomised protocol for
$GT_n$ using $t$ rounds of communication and having a complexity of
$O(n^{1/t} \log n)$. Hence, for a constant number of rounds, our
lower bound matches the upper bound to within
logarithmic factors. For one round quantum protocols, our result 
implies
an $\Omega(n)$ lower bound for $GT_n$ (which is optimal to within
constant factors), improving upon the 
previous $\Omega(n/\log n)$ lower bound of 
Klauck~\cite{klauck:ccsurvey}. No rounds versus communication tradeoff
for this problem, for more than one round,
was known earlier in the quantum setting.
If the number of rounds is unbounded, then there is a private
coin classical
randomised protocol for $GT_n$  using $O(\log n)$ rounds of 
communication and having a complexity of 
$O(\log n)$~\cite{nisan:threshold}. An
$\Omega(\log n)$ lower bound for the bounded error 
quantum communication
complexity of $GT_n$ (irrespective of the number of rounds)
follows from Kremer's
result~\cite{kremer:quantcc} that the bounded error quantum
communication complexity of a function is lower bounded (up to
constant factors) by the
logarithm of the one round (classical)
deterministic communication complexity.

\subsection{Our techniques}
The starting point of our work is the paper of Miltersen, Nisan, Safra
and Wigderson~\cite{miltersen:roundelim} showing lower 
bounds for randomised
cell probe schemes for predecessor. 
The crux of Miltersen et al.'s lower bound is the following
round elimination lemma for communication complexity. 

\bigskip

\noindent {\bf Fact (Round elimination lemma, 
\cite{miltersen:roundelim})}
{\em 
Let $f:\cX \times \cY \rightarrow \cZ$ be a function.
Let $\epsilon, \delta > 0$ be real numbers. Suppose that
$\delta \leq \epsilon^2 (100 \ln (8/\epsilon))^{-1}$.
Suppose the communication game $f^{(n),A}$ has a 
$[t;l_1,\ldots,l_t]^A$ public coin randomised protocol
with error less than $\delta$. Also suppose that
$n \geq 20 (l_1 \ln 2 + \ln 5) \epsilon^{-1}$.
Then there is a $[t-1;l_2,\ldots,l_t]^B$ public coin
randomised protocol for $f$ with error less than 
$\epsilon$.
}

\bigskip

\noindent Note that in the round elimination lemma 
of \cite{miltersen:roundelim}, the
dependence between $\delta$ and $\epsilon$ is quadratic.
In their paper, Miltersen et al. ask if their 
round elimination based
approach can be strengthened to obtain Beame 
and Fich's~\cite{beame:pred} lower bound.

Our first observation
is that if we can prove a stronger round elimination lemma in 
which $\delta$ and $\epsilon$ are related by a {\em small additive
term}, and the additive term is upper
bounded by $(\frac{l_1}{n})^{\Omega(1)}$,
then we can obtain the lower bound of Beame and Fich for
randomised cell probe schemes solving the predecessor problem.

Our next observation is that Klauck, Nayak, Ta-Shma and 
Zuckerman~\cite{klauck:interaction} have studied  rounds 
versus quantum communication tradeoffs
for the `tree pointer chasing' problem
using tools from quantum information theory. 
In fact, their
quantum lower bound for the `tree pointer chasing' problem is
better than its previously known classical 
lower bound~\cite{miltersen:roundelim}!
An important ingredient of their quantum lower bound was a
quantum information-theoretic result called 
{\em average encoding} theorem. This result says informally
that when messages carry very 
little information about the input, the average message is 
essentially 
as good as the individual messages. This result gives us a new way
of attacking the round elimination problem. The information-theoretic
round reduction
arguments in \cite{klauck:interaction} are 
average-case (under the uniform
distribution on the inputs) arguments, and do not immediately
give a worst-case result like the round elimination lemma. 
The information-theoretic arguments
have to be combined with 
Yao's minimax lemma~\cite{yao:minimax} (also used in the
proof of the round elimination lemma in 
\cite{miltersen:roundelim}) to prove
the strong round elimination lemma of this paper. 
The information-theoretic approach
brings out more clearly the
intuition behind round elimination, 
as opposed to the ad hoc combinatorial 
proof in \cite{miltersen:roundelim}.
We believe that this strong round
elimination lemma is an important technical contribution
of this paper. 

\subsection{Organisation of the paper}
We start with some preliminaries in the next 
section. Assuming the average encoding theorem, we prove an 
intermediate result in Section~\ref{sec:rndreduce} 
which allows us to reduce the number of rounds
of a communication protocol if the first message does not convey
much information about the sender's input. 
Proofs of the classical and quantum versions of
the average encoding
theorem can be found in the appendix for completeness.
Using the intermediate result, we
prove our strong round elimination lemma in
Section~\ref{sec:roundelim}. 
Sections~\ref{sec:rndreduce} and \ref{sec:roundelim} 
each have two subsections: the first
one treats the classical version of the results and 
the second one treats the quantum version.
Using the strong round elimination lemma, we prove the optimal
lower bound for the predecessor problem in Section~\ref{sec:predlb}. 
The rounds versus communication tradeoff for the `greater-than' 
problem is sketched in Section~\ref{sec:gt}. 
We finally conclude
mentioning some open problems in Section~\ref{sec:conclusion}.

\section{Preliminaries}

\subsection{The address-only quantum cell probe model}
\label{subsec:quantcell} 
A quantum {\em $(s,w,t)$ cell probe scheme} for a 
static data structure 
problem $f: D \times Q \rightarrow A$
has two components: a classical deterministic 
{\em storage scheme} that 
stores the data $d \in D$ in a table $T[d]$ using
$s$ cells each containing $w$ bits, and a quantum {\em query scheme}
that answers queries by `quantumly probing a cell at a time' 
$t$ times.
Formally speaking, the table $T[d]$ is made available to the
query algorithm in the form of an oracle unitary transform $O_d$. To
define $O_d$ formally, we represent the basis states of the query
algorithm as
$\ket{j,b,z}$, where $j \in [s-1]$ is a binary string of
length $\log s$, $b$ is a binary string of length $w$, and 
$z$ is a binary string of
some fixed length. Here, 
$j$ denotes the address of a cell in the table $T[d]$,
$b$ denotes the qubits which will hold the contents of a cell
and $z$ stands for the rest of the qubits (`work qubits')
in the query algorithm.
$O_d$ maps $\ket{j,b,z}$ to $\ket{j, b \oplus T[d]_j, z}$, where
$T[d]_j$ is a bit string of length $w$ and 
denotes the contents of the $j$th cell in $T[d]$.
A quantum query scheme with $t$ probes is just a sequence of
unitary transformations
\begin{displaymath}
U_{0} \rightarrow O_d \rightarrow U_{1} \rightarrow O_d
      \rightarrow \ldots U_{t-1}
      \rightarrow O_d \rightarrow U_{t}
\end{displaymath}
where $U_{j}$'s are arbitrary unitary transformations that do not
depend on $d$ ($U_{j}$'s represent the internal computations
of the query algorithm). For a query $q \in Q$, the
computation starts in a computational basis state
$\ket{q}\ket{0}$, where we
assume that the ancilla qubits are initially in the basis state
$\ket{0}$. Then we apply in succession,
the operators $U_0, O_d, U_1, \ldots, U_{t-1}, O_d, U_t$, 
and measure the final state.
The answer consists of the values on some of the
output wires of the circuit. 
We say that the scheme has worst case error probability less than
$\epsilon$ if 
the answer is equal to $f(d,q)$, for
every $(d,q) \in D \times Q$, with
probability greater than $1 - \epsilon$. 
The term `bounded error quantum scheme' means that 
$\epsilon = 1/3$.

We now formally define the {\em address-only quantum cell 
probe model}.
Here the storage scheme is classical deterministic as before,
but the query scheme is restricted to be `address-only quantum'.
This means that the state vector before a query to the oracle $O_d$
is always a {\em tensor product} of a state vector 
on the address and work qubits (the $(j,z)$ part in 
$(j,b,z)$ above), and a state 
vector on the data qubits (the $b$ part in
$(j,b,z)$ above). The state vector on the data qubits before a
query to the oracle $O_d$ is {\em independent of the 
query element $q$ and the data $d$} but
can vary with the probe number. Intuitively, we are only
making use of quantum parallelism over the address lines of
a query. This
mode of querying a table subsumes classical (deterministic or
randomised) querying, and also many 
non-trivial
quantum algorithms like Grover's algorithm~\cite{grover:search},
Farhi et al.'s algorithm~\cite{farhi:search}, 
H{\o}yer et al.'s algorithm~\cite{hoyer:search} etc. 
satisfy the `address-only' condition.
For classical (deterministic or randomised)
querying, the state vector on the data qubits is 
$\ket{0}$, independent of the probe number. 
For Grover's algorithm and Farhi et al.'s algorithm,
the state vector on the data qubit is
$(\ket{0} - \ket{1}) / \sqrt{2}$, independent of the probe number.
For H{\o}yer et al.'s algorithm, 
the state vector on the data qubit is 
$\ket{0}$ for some probe numbers, and 
$(\ket{0} - \ket{1}) / \sqrt{2}$ for the other probe numbers.

\subsection{Quantum communication protocols}
\label{subsec:quantcomm} 
In this paper, we adopt the `interacting unitary quantum circuits' 
definition of quantum
communication protocols of Yao~\cite{yao:quantcc}.
Thus, Alice and Bob send a certain number of fixed length
messages to each other, and the number and length of these messages
is independent of their inputs.
If Alice's and Bob's inputs are in computational basis
states, the global state of all the qubits of Alice and Bob is pure
at all times during the execution of the protocol.
Measurements are not allowed during the execution of 
the protocol. At the end of the protocol, the last recipient of
a message make a von Neumann measurement in the computational basis
of certain qubits (the `answer qubits')
in her possession in order to determine the answer
of the protocol. The choice of `answer qubits' is independent of
Alice's and Bob's inputs. 

We require that Alice and Bob
make a secure copy of their inputs before beginning the quantum
communication protocol.
This is possible since the inputs to Alice and Bob are in 
computational basis states e.g. CNOT gates can be used for this
purpose. Thus, without loss of generality,
the input qubits of Alice and Bob are never sent
as messages, their state remains unchanged throughout the protocol,
and they are never
measured i.e. some work qubits are measured to determine the result of
the protocol. 
We call such quantum protocols {\em secure}
and will assume
henceforth that all our quantum protocols are secure. 

We now define the concept of a {\em safe} 
quantum communication
protocol, which will be used in the statement of the 
quantum round elimination lemma.
\begin{definition}[Safe quantum protocol] 
A $[t;c;l_1,\ldots,l_t]^A$ ($[t;c;l_1,\ldots,l_t]^B$) safe quantum 
communication
protocol is a secure quantum protocol where Alice (Bob) starts
the communication, the first message is $l_1 + c$ qubits long,
the  $i$th message, for $i \ge 2$, is $l_i$ qubits long, and
the communication goes on for $t$ rounds. We think of the
first message as having two parts: the `main part' which is
$l_1$ qubits long, and the `safe overhead part' which is $c$ qubits
long. The density matrix of the `safe overhead' is independent of the 
inputs to Alice and Bob. 
\end{definition}

\paragraph{Remarks:} \  \\
1. \ \ The safe overhead is nothing but a way to send 
input independent
prior entanglement from Alice to Bob. The reason we use 
this notation is that we will later define safe public coin
quantum protocols where there will be two kinds of input
independent prior entanglement, the first being the one provided
by the safe overhead and the second being the one proved by the
public coin. \\
2. \ \ The reason for defining the concept of a safe overhead, 
intuitively speaking, is as follows.
The communication games arising from data structure problems
often have an asymmetry between the message lengths of Alice and Bob.
This asymmetry is crucial to prove lower bounds on the number
of rounds of communication.
In the previous quantum round reduction arguments (e.g. those
of Klauck et al.~\cite{klauck:interaction}), the complexity
of the first message in the protocol increases quickly as the
number of rounds is reduced and the asymmetry gets lost.
This leads to a problem where the
first message soon gets big enough to potentially
convey substantial information about the input of one player to the
other, destroying any hope of proving strong lower bounds on the
number of rounds.
The concept of a safe protocol allows us to get around this problem.
We show through a careful
quantum information theoretic analysis of the round reduction process,
that in a safe protocol,
though the complexity of the first message increases a lot, this
increase is confined to the safe overhead and so, the
information content does not increase much. This is the key property
which allows us to prove a round elimination lemma for safe
quantum protocols.

In this paper we will deal with quantum protocols with {\em public
coins}. Intuitively, a public coin quantum protocol is 
a probability distribution over finitely many ({\em coinless})
quantum protocols. We shall
henceforth call the standard definition of a quantum protocol
without prior entanglement as {\em coinless}. Our definition
is similar to the classical scenario, where a randomised protocol
with public coins is a probability distribution over finitely
many deterministic
protocols. We note however, that our definition of a 
public coin quantum
protocol is {\em not} the same as that of a 
quantum protocol with prior
entanglement, which has been studied previously 
(see e.g.~\cite{cleve:ip}). Our definition is weaker, in that it
does not allow the unitary transformations of Alice and Bob to
alter the `public coin'.
\begin{definition}[Public coin quantum protocol]
In a quantum protocol with a public coin, there is, before the
start of the protocol, a 
quantum state called a {\em public coin}, of the form
$\sum_c \sqrt{p_c} \ket{c}_A \ket{c}_B$, where the
subscripts denote ownership of qubits by Alice and Bob, 
$p_c$ are finitely many non-negative real
numbers and $\sum_c p_c = 1$. 
Alice and Bob make (entangled) copies of their respective 
halves of the
public coin using {\sc CNOT} gates before commencing the
protocol. The unitary transformations of Alice and Bob during the
protocol do not touch the public coin.
The public coin is never measured, nor is it ever sent as a message.
\end{definition}
Hence, one can
think of the public coin quantum protocol to be a probability
distribution, with probability $p_c$, over finitely many
coinless quantum protocols
indexed by the coin basis states $\ket{c}$. A {\em safe public coin}
quantum protocol is similarly defined as
a probability distribution over finitely many safe
coinless quantum protocols.

\paragraph{Remarks:} \  \\ 
1.\ \ We need to define public coin quantum protocols in order to 
make use of the harder direction of Yao's minimax 
lemma~\cite{yao:minimax}. The minimax lemma
is the main tool which allows us to convert `average case'
round reduction arguments to `worst case' arguments. We need `worst
case' round reduction arguments in proving lower 
bounds for the rounds
complexity of communication games arising from data
structure problems. This is because many of these lower
bound proofs
use some notion of ``self-reducibility'', arising from the 
original data
structure problem, which fails to hold in the `average case' but holds
for the `worst case'. The quantum round reduction arguments of 
Klauck et al.~\cite{klauck:interaction}
are `average case' arguments, and this is 
one of the reasons why they do not suffice to prove lower bounds
for the rounds complexity of communication games arising from data
structure problems. \\
2.\ \  For $[t;c;l_1,\ldots,l_t]^A$ safe quantum protocols computing
a function $f$, Yao's minimax lemma says that
the infimum (worst-case) error of a public coin protocol is equal
to the supremum over all input probability distributions of
the infimum distributional error of a coinless protocol.

\subsection{Predecessor searching and communication complexity}
\label{subsec:mnswpredlb}
We first describe the connection between the address-only
quantum cell probe
complexity of a static data structure problem and the quantum
communication complexity of an associated communication game.
Let $f: D \times Q \rightarrow A$ be a static data structure problem.
Consider a two-party
communication problem where Alice is given a query $q \in Q$,
Bob is given a datum $d \in D$, and they have to communicate and
find out the answer $f(d,q)$. We have the following lemma, which
is a quantum analogue of a lemma of Miltersen~\cite{miltersen:union}
relating cell probe complexity to communication complexity in
the classical setting.
\begin{lemma}
\label{lem:queryandcc}
Suppose we have an $(s,w,t)$ quantum cell 
probe solution to a static data structure problem 
$f: D \times Q \rightarrow A$.
Then we have a
$(2t, 0, \log s + w, \log s + w)^A$ safe coinless quantum
protocol for the corresponding
communication problem. If the query scheme is address-only,
we also have a $(2t, 0, \log s, \log s + w)^A$ safe coinless 
quantum protocol for the corresponding communication problem.
The error probability of the communication protocol is the same as
that of the cell probe scheme.
\end{lemma}
\begin{proof} 
Given a quantum $(s,w,t)$ cell
probe solution to the static data structure problem $f$,
we can get a 
$(2t, 0, \log s + w, \log s + w)^A$ safe coinless quantum
protocol for the corresponding
communication problem by just
simulating the cell probe solution. If in addition,
the query scheme is address-only, the messages from Alice to
Bob need consist only of the `address' part. This can be
seen as follows. Let the state vector of the data qubits before
the $i$th query be $\ket{\theta_i}$. $\ket{\theta_i}$ is independent
of the query element and the stored data. 
Bob keeps $t$ special ancilla registers in states 
$\ket{\theta_i}, 1 \leq i \leq t$ at the start of the protocol $P$.
These special ancilla registers are in tensor with the rest of the
qubits of Alice and Bob at the start of $P$.
Protocol $P$ simulates the cell probe solution, but with the following
modification. To simulate the $i$th query of the cell probe
solution, Alice prepares her `address' and `data' qubits as in the
query scheme, but sends the `address' qubits only. Bob treats those
`address' qubits together with $\ket{\theta_i}$ in the $i$th
special ancilla register as Alice's query, and performs the oracle
table transformation on them. He then sends these qubits (both
the `address' as well as the $i$th special register qubits) to Alice.
Alice exchanges the contents of the $i$th special register with
her `data' qubits (i.e. exchanges the basis states), and proceeds
with the simulation of the query scheme. This gives us a 
$(2t, 0, \log s, \log s + w)^A$ safe coinless 
quantum protocol with the same error probability as
that of the cell probe query scheme.
\end{proof}

\paragraph{Remark:}
In many natural data structure problems $\log s$ is much smaller than
$w$ and thus, in the address-only quantum case, we get a 
$(2t, 0, \log s, O(w))^A$ safe protocol. In the classical
setting of \cite{miltersen:union}, one gets a 
$(2t, 0, \log s, w)^A$ protocol. 
This asymmetry in message lengths is crucial
in proving non-trivial lower bounds on $t$.
The concept of a safe quantum protocol helps us in exploiting
this asymmetry.

We now recall some facts about the connection between 
cell probe schemes for predecessor and communication
complexity of rank parity
from \cite{miltersen:roundelim}. We give proof sketches of these
facts  for completeness.

\begin{definition}
A $(t,a,b)^A$ ($(t,a,b)^B$) classical communication
protocol is a 
$[t;l_1,\ldots,l_t]^A$ ($[t;l_1,\ldots,l_t]^B$)
classical protocol, where assuming
Alice (Bob) starts, $l_i = a$ for $i$ odd and $l_i = b$ for
$i$ even ($l_i = b$ for $i$ odd and $l_i = a$ for $i$ even).
A $(t,c,a,b)^A$ ($(t,c,a,b)^B$) quantum communication
protocol is a 
$[t;c;l_1,\ldots,l_t]^A$ ($[t;c;l_1,\ldots,l_t]^B$)
safe quantum protocol, where assuming
Alice (Bob) starts, $l_i = a$ for $i$ odd and $l_i = b$ for
$i$ even ($l_i = b$ for $i$ odd and $l_i = a$ for $i$ even).
\end{definition}

\begin{definition}[Rank parity]
In the {\em rank parity} communication
game $\PAR_{p,q}$,
Alice is given a bit string $x$ of
length $p$, Bob is given a set $S$ of bit strings of length $p$,
$|S| \leq q$, and they have to communicate and decide
whether the rank of $x$ in $S$ (treating the bit 
strings as integers) is odd or even. 
By the rank of $x$ in $S$, we mean the cardinality of the
set $\{y \in S \mid y \leq x \}$.
In the communication game $\PAR^{(k),A}_{p,q}$,
Alice is given $k$ bit strings $x_1,\ldots,x_k$ each of 
length $p$, Bob is given a set $S$ of bit strings of length $p$,
$|S| \leq q$, an index $i \in [k]$, 
and copies of $x_1,\ldots,x_{i-1}$;
they have to communicate and decide 
whether the rank of $x_i$ in $S$ is odd or even.
In the communication game $\PAR^{(k),B}_{p,q}$,
Alice is given a bit string $x$ of 
length $p$ and an index $i \in [k]$, Bob is given $k$ sets
$S_1,\ldots,S_k$ of bit strings of length $p$,
$|S_j| \leq q, 1 \leq j \leq k$;
they have to communicate and decide 
whether the rank of $x$ in $S_i$ is odd or even.
\end{definition}

\begin{fact}
\label{fact:pred2rank}
Let $m$ be a positive integer such that $m$ is a power of $2$.
Suppose that there is a $(n^{O(1)}, (\log m)^{O(1)}, t)$ 
randomised (address-only quantum) cell 
probe scheme for the $(m, n)$-static predecessor problem.
Then the rank parity communication game $\PAR_{\log m, n}$ has a
$\left(2t+O(1), O(\log n), (\log m)^{O(1)} \right)^A$ 
private coin randomised (safe coinless quantum) protocol.
The error probability of the communication protocol is the same as
that of the cell probe scheme.
\end{fact}
\begin{proof}
Consider the static rank parity data structure problem where 
the storage scheme has to store a set $S \subseteq [m]$, $|S| \leq n$,
and the query scheme, given a query $x \in [m]$, has to decide whether
the rank of $x$ in $S$ is odd or even.
Fredman, Koml\'{o}s and Szemer\'{e}di~\cite{fredman:hashing}
have shown the existence of two-level 
perfect hash tables containing, for each member $y$ of the
stored subset $S$, $y$'s rank in $S$, and
using $O(n)$ cells of word size $O(\log m)$ and requiring
only $O(1)$ deterministic cell probes.
Combining a $(n^{O(1)}, (\log m)^{O(1)}, t)$ 
cell probe solution to the static predecessor problem with such
a perfect hash table gives us a 
$(n^{O(1)} + O(n), \max((\log m)^{O(1)}, O(\log m)), t + O(1))$ 
cell probe solution to the static rank parity problem.
The error probability of the cell probe scheme for the rank
parity problem is the same as the error probability of the cell
probe scheme for the predecessor problem.
Converting the cell probe scheme for the rank parity problem
into a communication protocol by \cite{miltersen:union} 
(by Lemma~\ref{fact:pred2rank}), we get a
$(2t+O(1), O(\log n), (\log m)^{O(1)})^A$ private coin 
randomised (safe coinless quantum) protocol for the rank
parity communication game $\PAR_{\log m, n}$. 
The error probability of the communication protocol is the same as
that of the cell probe scheme for the predecessor problem.
\end{proof}

\begin{fact}
\label{fact:rankred1}
Let $k, p$ be positive integers such that $k \mid p$.
A communication protocol with Alice starting for $\PAR_{p,q}$
gives us a communication protocol with 
Alice starting for $\PAR^{(k),A}_{p/k,q}$ with the 
same message complexity, number of rounds 
and error probability.
\end{fact}
\begin{proof}
Consider the problem $\PAR^{(k),A}_{p/k,q}$.
Alice, who is given $x_1,\ldots,x_k$, computes the concatenation
$\widehat{x} \defeq x_1 \cdot x_2 \cdots x_k$. Bob, who is given
$S$, $i$ and $x_1,\ldots,x_{i-1}$, computes 
\begin{displaymath}
\widehat{S} \defeq 
            \left\{x_1 \cdot x_2 \cdots x_{i-1} 
                       \cdot y \cdot 0^{p(1 - i/k)}:  y \in S
            \right\}.
\end{displaymath}
After this, Alice and Bob run the protocol for $\PAR_{p,q}$ on inputs
$\widehat{x}$, $\widehat{S}$ to solve the 
problem $\PAR^{(k),A}_{p/k,q}$.
\end{proof}

\begin{fact}
\label{fact:rankred2}
Let $k, q$ be positive integers such that 
$k \mid q$ and $k$ is a power of $2$.
A communication protocol with Bob starting for $\PAR_{p,q}$
gives us a communication protocol with Bob starting for 
$\PAR^{(k),B}_{p - \log k - 1,q/k}$ with the same message complexity, 
number of rounds and error probability.
\end{fact}
\begin{proof}
Consider the problem $\PAR^{(k),B}_{p - \log k - 1,q/k}$.
Alice, who is given $x$ and $i$, 
computes the concatenation $\hat{x} \defeq (i-1) \cdot 0 \cdot x$.
Bob, who is given $S_1,\ldots,S_k$, computes the sets 
$S'_1,\ldots,S'_k$ where
\begin{displaymath}
S'_j \defeq  
       \left\{ \begin{array}{l}
                 \left\{(j-1) \cdot 0 \cdot y: y \in S_j \right\}\\
                 ~~~~~~~~~~~~~~~~~~~~ \mbox{if $|S_j|$ is even}, \\
                 \left\{(j-1) \cdot 0 \cdot y: y \in S_j \right\} 
                  \bigcup \left\{(j-1) \cdot 1^{p - \log k} \right\}\\
                 ~~~~~~~~~~~~~~~~~~~~ \mbox{if $|S_j|$ is odd}.
               \end{array}
       \right.
\end{displaymath}
Above, the integers $(i-1), (j-1)$ are to be thought of as 
bit strings of length $\log k$.
Bob also computes $\hat{S} \defeq \bigcup_{j=1}^k S'_j$.
Alice and Bob then run the protocol for $\PAR_{p,q}$ on inputs
$\hat{x}$, $\hat{S}$ to solve the problem 
$\PAR^{(k),B}_{p - \log k - 1,q/k}$.
\end{proof}

\subsection{Some classical information theoretic facts}
\label{sec:classicalinfoprelim}
In this subsection, we discuss some classical
information theoretic facts
which will be used in the proof of our improved classical
round elimination lemma.
For a good account of classical information theory, see 
the book by Cover and Thomas~\cite{cover:infotheory}.

In this paper, all random variables have finite range and all
sample spaces have finite cardinality.
Let $X,Y,Z$ be random variables with some joint distribution.
The {\em Shannon entropy} of $X$ is defined as
$H(X) \defeq -\sum_x \Pr[X=x] \log \Pr[X=x]$. 
The {\em mutual information} of $X$ and $Y$ is defined
as $I(X:Y) \defeq H(X) + H(Y) - H(XY)$. If the range of $X$
has cardinality at most $d$, $I(X:Y) \leq \log d$. 
$I(X:Y) = 0$ iff $X$ and $Y$ are independent.
Let $I((X:Y) \mid Z = z)$ denote the mutual information of 
$X$ and $Y$ conditioned on the event $Z=z$.

The next fact follows easily from the definitions.
\begin{fact}
\label{fact:classicalaveraging}
Let $X,Y,Z$ be random variables with some joint distribution.
Then, 
\begin{enumerate}
\item[(a)] $I(XY:Z) = I(X:Z) + I(Y:ZX) - I(X:Y)$. 
           In particular, if $X, Y$ are independent 
           $I(XY:Z) = I(X:Z) + I(Y:ZX)$. 
\item[(b)] $\displaystyle 
            I(Y:ZX) = I(X:Y) + \E_{X} [I((Y:Z) \mid X=x)]$, where
           is expectation is over the marginal distribution of $X$. 
\end{enumerate}
\end{fact}

We use total variation distance to quantify the distance between
two probability distributions.
\begin{definition}[Total variation distance]
Let $P,Q$ be probability distributions on the same 
sample space $\Omega$. 
The {\em total variation distance} (also known as the
$\ell_1$-distance) between
$P$ and $Q$, denoted by $\ellone{P - Q}$, is defined as
$\displaystyle \ellone{P - Q} \defeq \sum_{x \in \Omega} |P(x)-Q(x)|$.
\end{definition}

We will need the {\em average encoding} theorem of
Klauck, Nayak, Ta-Shma and Zuckerman~\cite{klauck:interaction}. 
Klauck et al. actually prove a quantum version of this
theorem in their paper, but we will use the classical version
in the proof of our round elimination lemma.
Intuitively speaking, the theorem says that if the mutual information 
between a random variable and its randomised encoding is small, 
then the probability distributions on the code words 
for various values of the random variable
are close to the average probability distribution on the code words.
\begin{fact}[Average encoding theorem, \cite{klauck:interaction}]
\label{fact:averageencodingclassical}
Let $X, M$ be correlated 
random variables. Let $p_x$ denote the (marginal) probability
that $X=x$, and $\Pi^x$ denote the conditional distribution of $M$
given that $X=x$.
Let $\Pi$ denote the (marginal) probability distribution of $M$ i.e.
$\Pi = \sum_x p_x \Pi^x$. 
Then,
\begin{displaymath}
\sum_x p_x \ellone{\Pi^x - \Pi} \leq \sqrt{(2 \ln 2) I(X:M)}.
\end{displaymath}
\end{fact}
A self-contained classical
proof, without using quantum information theory,
of this fact can be found in the appendix for completeness.

\subsection{Some quantum information theoretic facts}
\label{sec:quantinfoprelim}
In this subsection, we discuss some quantum
information theoretic facts
which will be used in the proof of our quantum
round elimination lemma.
For a good account of quantum information theory, see 
the book by Nielsen and Chuang~\cite{nielsen:quant}.

In this paper, all quantum systems, Hilbert spaces and superoperators
are finite dimensional.
Let $X,Y$ be quantum systems with some joint density matrix 
$\rho_{XY}$.
If $\rho_X$ is the reduced density matrix of $X$,
the {\em von Neumann entropy} of $X$ is defined as
$S(X) \defeq -\Tr \rho_X \log \rho_X$. 
The {\em mutual information} of $X$ and $Y$ is defined
as $I(X:Y) \defeq S(X) + S(Y) - S(XY)$. If $X$
is at most $d$-dimensional, $I(X:Y) \leq 2 \log d$. 
$I(X:Y) = 0$ iff $X$ and $Y$ are independent i.e.
$\rho_{XY} = \rho_X \otimes \rho_Y$.
Suppose $X,Y,Z$ are quantum systems with some joint density matrix
where $Z$ is a classical random variable i.e. the reduced density
matrix of $Z$ is diagonal in the computational basis.
Let $I((X:Y) \mid Z = z)$ denote the mutual information of 
$X$ and $Y$ conditioned on the event $Z=z$.

The next fact follows easily from the definitions.
\begin{fact}
\label{fact:quantaveraging}
Let $X,Y,Z,W$ be quantum systems with some joint density matrix,
where $X$ and $Y$ are classical random variables. 
Then, 
\begin{enumerate}
\item[(a)] $I(XY:Z) = I(X:Z) + I(Y:ZX) - I(X:Y)$. 
           In particular, if $X, Y$ are independent 
           $I(XY:Z) = I(X:Z) + I(Y:ZX)$. 
\item[(b)] $\displaystyle 
            I(Y:ZX) = I(X:Y) + \E_{X} [I((Y:Z) \mid X=x)]$, where
           is expectation is over the marginal distribution of $X$. 
\item[(c)] Suppose $W$ is independent of $X$ and $Z$ is supported
           on $m$ qubits. Then, $I(X:ZW) \leq 2m$.
\end{enumerate}
\end{fact}

\paragraph{Remarks:} \ \\
1. \ \ Fact~\ref{fact:quantaveraging}(c) is the key observation 
allowing us to
``ignore'' the size of the ``safe'' overhead $W$ 
in quantum round elimination applications.
In these applications, the complexity of the first message
in the protocol increases quickly, but the blow up is confined to
the ``safe'' overhead. Earlier round reduction arguments were unable
to handle this large blow up in the complexity of the first message.\\
2. \ \ In Fact~\ref{fact:quantaveraging}(c), if $Z$ is a 
classical random variable, we get the improved
inequality $I(X:ZW) \leq m$.

We use trace distance to quantify the distance between
two density matrices.
\begin{definition}[Trace distance]
Let $\rho,\sigma$ be density matrices in the same Hilbert space.
The {\em trace distance} between
$\rho$ and $\sigma$, denoted by $\ellone{\rho - \sigma}$, 
is defined as
$\ellone{\rho - \sigma} \defeq 
 \Tr \sqrt{(\rho - \sigma)^\dagger (\rho - \sigma)}$.
\end{definition}
If $\rho$ is a density matrix in a Hilbert space $\cH$ and
$\cM$ is a general measurement i.e. a POVM on $\cH$, let
$\cM \rho$ denote the probability distribution on
the (classical) outcomes of $\cM$ got by performing measurement
$\cM$ on $\rho$. 
The importance of the trace distance as a metric on density 
matrices stems from the following fundamental fact 
(see e.g.~\cite{aharonov:mixed}).
\begin{fact}
\label{fact:l1tracedist}
Let $\rho_1,\rho_2$ be two density matrices in the same Hilbert
space $\cH$. Let $\cM$ be a POVM on $\cH$. Then,
$\ellone{\cM\rho_1 - \cM\rho_2} \leq \ellone{\rho_1 - \rho_2}$.
\end{fact}

Let $\cH, \cK$ be disjoint Hilbert spaces such that 
$\dim(\cK) \geq \dim(\cH)$. Let $\rho$ be a density matrix in
$\cH$ and $\ket{\psi}$ be a pure state in $\cH \otimes \cK$.
$\ket{\psi}$ is said to be a {\em purification} of $\rho$
if $\Tr_{\cK} \ketbra{\psi} = \rho$. We will require the following
basic fact about two purifications of the same density matrix.
\begin{fact}
\label{fact:localtransition}
Let $\cH, \cK$ be disjoint Hilbert spaces such that 
$\dim(\cK) \geq \dim(\cH)$. Let $\rho$ be a density matrix in
$\cH$ and $\ket{\psi}, \ket{\phi}$ be two purifications of
$\rho$ in $\cH \otimes \cK$. Then there is a local
unitary transformation $U$ on $\cK$ such that
$\ket{\psi} = (I \otimes U) \ket{\phi}$, where $I$ is the
identity operator on $\cH$.
\end{fact}

We now state an improved version of the quantum 
{\em average encoding} theorem of
Klauck, Nayak, Ta-Shma and Zuckerman~\cite{klauck:interaction}. 
This improved
version follows from a direct connection between fidelity and
relative entropy described in \cite{dacunha:BgeqS}, and was 
also independently observed by Klauck (private communication).
Intuitively speaking, the theorem says that if the mutual information 
between a random variable and its quantum encoding is small, 
then given any purifications of the quantum code words
for various values of the random variable,
one can find purifications of the average code word that are close
to the respective purifications of the quantum code words.
\begin{fact}
\label{fact:averageencodingquant}
Let $X, M$ be quantum systems with some joint density matrix.
Let $X$ be a classical random variable
and $p_x$ denote the (marginal) probability
that $X=x$. Let $\cH$ denote the Hilbert space of $M$
and $\rho^x$ denote the conditional density matrix
of $M$ given that $X=x$. 
Let $\rho$ denote the reduced density matrix of $M$. Note that
$\rho = \sum_x p_x \rho^x$. 
Let $\cK$ denote a disjoint 
Hilbert space such that $\dim(\cK) \geq \dim(\cH)$.
Let $\ket{\psi^x}$ be purifications of $\rho^x$ in
$\cH \otimes \cK$. Then there exist purifications $\ket{\phi^x}$
of $\rho$ in $\cH \otimes \cK$ such that 
\begin{displaymath}
\sum_x p_x \ellone{\ketbra{\psi^x} - \ketbra{\phi^x}} \leq 
\sqrt{(4 \ln 2) I(X:M)}.
\end{displaymath}
\end{fact}
A proof of this fact can be found in the appendix
for completeness.

\section{Reducing the number of rounds}
\label{sec:rndreduce}
In this section, we prove an intermediate result
which will be required to prove our strong round elimination lemmas. 
In Section~\ref{subsec:classicalrndreduce} we prove the
intermediate result in its classical version 
(Lemma~\ref{lem:classicalroundreduce}), 
whereas in Section~\ref{subsec:quantrndreduce} we prove the
intermediate result in its quantum version
(Lemma~\ref{lem:quantroundreduce}).
The proof of Lemma~\ref{lem:classicalroundreduce}
is similar to the proof of Lemma~4.4 in 
\cite{klauck:interaction} (see also \cite{maneva:interaction}), 
but much simpler since we are in the classical setting. 
The proof of Lemma~\ref{lem:quantroundreduce}
is a slight refinement of the proof of Lemma~4.4 in 
~\cite{klauck:interaction}, and is included here for completeness.
Intuitively speaking, the intermediate result
says that if the first message
in a communication protocol
carries little information about the sender's
input, under some probability
distribution on Alice's and Bob's inputs, 
then it can be eliminated, giving rise
to a protocol where the other player starts, with one less 
round of communication, smaller message complexity, and with similar
average error probability with respect to the same probability
distribution on Alice's and Bob's inputs. 

Consider a communication protocol $\cP$ computing
a function $f: \cX \times \cY \rightarrow \cZ$.
For an input $(x,y) \in \cX \times \cY$, we define the error 
$\epsilon^{\cP}_{x,y}$ of 
$\cP$ on $(x,y)$ to be the probability 
that the result of $\cP$ on input $(x,y)$ is not 
equal to $f(x,y)$. For
a protocol $\cP$, given a probability distribution
$D$ on $\cX \times \cY$, we define the 
average error 
$\epsilon^{\cP}_D$ of $\cP$ with respect to $D$ as the expectation
over $D$ of the error of $\cP$ on inputs $(x,y) \in \cX \times \cY$. 
We define $\epsilon^\cP$ to be the maximum error of $\cP$ on inputs 
$(x,y) \in \cX \times \cY$ i.e. 
$\epsilon^\cP$ is the error of protocol $\cP$.

\subsection{Classical}
\label{subsec:classicalrndreduce}
\begin{lemma}
\label{lem:classicalroundreduce}
Suppose $f: \cX \times \cY \rightarrow \cZ$ is a function. Let $D$ be
a probability distribution on $\cX \times \cY$, and
$\cP$ be a $[t;l_1,\ldots,l_t]^A$ private coin
randomised protocol for $f$. 
Let $X$ stand for the classical random variable denoting 
Alice's input,
$M$ stand for the random variable denoting the
first message of Alice in $\cP$, and 
$I(X:M)$ denote the mutual information between $X$ and $M$ when
the inputs to $\cP$ are distributed according to $D$. 
Then there exists a $[t-1;l_2,\ldots,l_t]^B$ 
deterministic protocol $\cQ$ for $f$, such that
$\epsilon^\cQ_D \leq \epsilon^\cP_D + (1/2) ((2 \ln 2) I(X:M))^{1/2}$.
\end{lemma}
\begin{proof}
We first give an overview of the idea of the proof before
getting down to the details. The proof proceeds in stages. 

\paragraph{Idea of Stage 1:} Starting from protocol $\cP$, we 
construct a 
$[t;l_1,\ldots,l_t]^A$ private coin protocol $\cP'$
where the first message
is independent of Alice's input, and $\epsilon^{\cP'}_D 
\leq \epsilon^{\cP}_D + (1/2) ((2 \ln 2) I(X:M))^{1/2}$.
The important idea here is to first generate Alice's 
message using a new private coin without `looking' at her input, 
and after that,
to adjust Alice's old private coin in a suitable manner so as to
be consistent with her message and input. 

\paragraph{Idea of Stage 2:} Since the first message of 
$\cP'$ is independent
of Alice's input, Bob can generate it himself. 
Doing this and setting coin tosses appropriately gives us a 
$[t-1;l_2,\ldots,l_t]^B$ deterministic protocol
$\cQ$ for $f$ such that 
$\epsilon^\cQ_D \leq \epsilon^{\cP'}_D 
   \leq \epsilon^\cP_D + (1/2) ((2 \ln 2) I(X:M))^{1/2}$.

\bigskip

We now give the details of the proof.
Let $\Pi^x$ be the probability distribution of the first message 
$M$ of protocol $\cP$ when Alice's input $X=x$. 
Define $\Pi \defeq \sum_x d_x \Pi^x$, where
$d_x$ is the marginal probability of $X=x$ under distribution $D$.
$\Pi$ is the probability distribution of the average first
message of $\cP$ under distribution $D$.
For $x \in \cX$ and an instance $m$ of the first message of Alice, let 
$q^{xm}_r$ denote the conditional probability that the private
coin toss of Alice results in $r$, given that Alice's input is $x$
and her first message in $\cP$ is $m$. 
If in $\cP$ message $m$ cannot occur when Alice's 
input is $x$, then we
define $q^{xm}_r \defeq 0$.
Let $\pi_m^x$ 
denote the probability that the first message of 
Alice in $\cP$ is $m$,
given that her input is $x$. Let $\pi_m$ denote the 
probability that the first message of Alice in $\cP$ is $m$, when
Alice's and Bob's inputs are distributed according to $D$.
Then, $\pi_m = \sum_x d_x \pi_m^x$.

\paragraph{Stage 1: }
We construct a $[t;l_1,\ldots,l_t]^A$ private coin 
randomised protocol $\cP'$ for $f$ with average error under 
distribution $D$
$\epsilon^{\cP'}_D \leq \epsilon^\cP_D + 
                        (1/2) ((2 \ln 2) I(X:M))^{1/2}$,
and where the probability distribution of
the first message is independent of the input to Alice.
We now describe the protocol $\cP'$.
Suppose Alice is given
$x \in \cX$ and Bob is given $y \in \cY$. Alice tosses a fresh private
coin to pick $m$ with probability $\pi_m$ and set
her old private coin to $r$ with probability $q^{xm}_r$. 
After this, Alice and Bob behave as in protocol $\cP$ (henceforth,
Alice ignores the new private coin which she had tossed to generate
her first message $m$). 
Hence in $\cP'$ the probability distribution of the first message
is independent of Alice's input. 

Let us now compare the situations in protocols
$\cP$ and $\cP'$ when Alice's input is $x$, Bob's input is $y$, 
Alice has finished tossing her private coins (both old and new),
but no communication has taken place as yet. 
In protocol $\cP$, the probability that Alice's private coin toss 
results in $r$ is $\sum_m \pi_m^x q^{xm}_r$.
In protocol $\cP'$, the probability that Alice's old private coin 
is set to $r$ is $\sum_m \pi_m q^{xm}_r$.
Thus, the total variation distance between the probability 
distributions on Alice's old private coin is
\begin{eqnarray*}
\sum_r \left|\sum_m q^{xm}_r (\pi_m^x - \pi_m)\right| 
& \leq & \sum_r \sum_m q^{xm}_r \left| \pi_m^x - \pi_m\right| \\
&   =  & \sum_m \left(\left| \pi_m^x - \pi_m \right| 
                            \sum_r q^{xm}_r  
                \right) \\
&   =  & \sum_m \left| \pi_m^x - \pi_m \right| \\
&   =  & \ellone{\Pi^x - \Pi}.
\end{eqnarray*}
Hence, the error probability of $\cP'$ on input $x,y$ 
$\epsilon^{\cP'}_{x,y} \leq \epsilon^\cP_{x,y} + 
                         (1/2) \ellone{\Pi^x - \Pi}$.
Let $d_{xy}$ be the probability that $(X,Y)=(x,y)$ under distribution
$D$. Then, the average error of $\cP'$ under distribution $D$
\begin{eqnarray*}
\epsilon^{\cP'}_D 
  &   =  & \sum_{x,y} d_{xy} \epsilon^{\cP'}_{x,y} \\
  & \leq & \sum_{x,y} d_{xy} \left(\epsilon^\cP_{x,y} + \frac{1}{2} 
                                   \ellone{\Pi^x - \Pi}\right) \\
  &   =  & \epsilon^\cP_D + 
           \frac{1}{2} \sum_x d_x \ellone{\Pi^x - \Pi}   \\
  & \leq & \epsilon^\cP_D + \frac{1}{2} ((2 \ln 2) I(X:M))^{1/2}.
\end{eqnarray*}
The last inequality follows from 
Fact~\ref{fact:averageencodingclassical}.

\paragraph{Stage 2:} 
We now construct our desired $[t-1;l_2,\ldots,l_t]^B$ 
deterministic protocol $\cQ$ for $f$ with
$\epsilon^\cQ_D \leq \epsilon^{\cP'}_D$. 
Suppose all the coin tosses 
of Alice and Bob in $\cP'$ are done publicly before any communication
takes place. Now there is no need for the first message from 
Alice to Bob, because Bob can reconstruct the message by looking
at the public coin tosses. 
This gives us a $[t-1;l_2,\ldots,l_t]^B$ public coin
protocol $\cQ'$ such that
$\epsilon^{\cQ'}_{x,y} = \epsilon^{\cP'}_{x,y}$ for
every $(x,y) \in \cX \times \cY$. By setting the public coin tosses of
$\cQ'$ to an appropriate value, we get a $[t-1;l_2,\ldots,l_t]^B$
deterministic protocol
$\cQ$ such that 
$\epsilon^\cQ_D \leq \epsilon^{\cQ'}_D = \epsilon^{\cP'}_D
   \leq \epsilon^\cP_D + (1/2) ((2 \ln 2) I(X:M))^{1/2}$. 

This completes the proof of Lemma~\ref{lem:classicalroundreduce}.
\end{proof}

\subsection{Quantum}
\label{subsec:quantrndreduce}
\begin{lemma}
\label{lem:quantroundreduce}
Suppose $f: \cX \times \cY \rightarrow \cZ$ is a function. Let $D$ be
a probability distribution on $\cX \times \cY$, and
$\cP$ be a $[t;c;l_1,\ldots,l_t]^A$ safe coinless
quantum protocol for $f$. 
Let $X$ stand for the classical random variable denoting 
Alice's input, $M$ denote the
first message of Alice in $\cP$, and 
$I(X:M)$ denote the mutual information between $X$ and $M$ when
the inputs to $\cP$ are distributed according to $D$. 
Then there exists a $[t-1;c+l_1;l_2,\ldots,l_t]^B$ 
safe coinless quantum protocol $\cQ$ for $f$, such that
$\epsilon^\cQ_D \leq \epsilon^\cP_D + (1/2) ((4 \ln 2) I(X:M))^{1/2}$.
\end{lemma}
\begin{proof}
We first give an overview of the plan of the proof, before
getting down to the details. The proof proceeds in stages.
Stage~1 of the quantum proof corresponds to
Stage~1 of the classical proof, and
Stages~2A and 2B of the quantum proof together correspond to
Stage~2 of the classical proof.

\paragraph{Idea of Stage 1:}
Starting from protocol $\cP$, we construct a 
$[t;c;l_1,\ldots,l_t]^A$ safe coinless quantum protocol $\cP'$
where the first message
is independent of Alice's input, and 
$\epsilon^{\cP'}_D \leq 
 \epsilon^{\cP}_D + (1/2)((4 \ln 2) I(X:M))^{1/2}$.
The important idea here is to generate a purification $\ket{\phi^x}$
of the average first 
message $\rho$ of Alice in $\cP$ that is close to
the purification $\ket{\psi^x}$ of the message $\rho^x$ of Alice
in $\cP$ when her input is $x$. The existence of such a purification
$\ket{\phi^x}$ is guaranteed by Fact~\ref{fact:averageencodingquant}.

\paragraph{Idea of Stage 2A:}
Since the first message of $\cP'$ is independent of
Alice's input (in fact its density matrix is $\rho$), 
Bob can generate it himself. This suffices if $\cP'$ is a one-round
protocol, and completes the proof of Lemma~\ref{lem:quantroundreduce}
for such protocols.
But if $\cP'$ has more than one round, it is also necessary for Bob
to achieve the correct entanglement between
Alice's work qubits and the first message i.e. it is necessary that
the joint state of Alice's work qubits and the qubits of the first
message be $\ket{\phi^x}$. 
Bob achieves this by first sending a safe message of $l_1+c$ qubits.
If Alice's input is $x$, she then applies a unitary transformation 
$V_x$ on her work qubits in order to make 
the joint state of her work qubits and the qubits of the first
message $\ket{\phi^x}$. 
The existence of such a $V_x$ follows from
Fact~\ref{fact:localtransition}.
Doing all this gives us a 
$[t+1;c+l_1;0,0,l_2,\ldots,l_t]^B$ safe coinless quantum
protocol $\cQ'$ for $f$ such that  
$\epsilon^{\cQ'}_{x,y} = \epsilon^{\cP'}_{x,y}$ for
every $(x,y) \in \cX \times \cY$. 

\paragraph{Idea of Stage 2B:}
Since the first message of Alice in $\cQ'$ is zero qubits
long, Bob can concatenate his first two messages, giving
us a $[t-1;c+l_1;l_2,\ldots,l_t]^B$ safe coinless quantum
protocol $\cQ$ for $f$ such that
$\epsilon^{\cQ}_{x,y} = \epsilon^{\cQ'}_{x,y}$ for
every $(x,y) \in \cX \times \cY$. The technical reason
behind this is that unitary transformations on disjoint
sets of qubits commute. 
Note that
$\epsilon^{\cQ}_D = \epsilon^{\cQ'}_D = \epsilon^{\cP'}_D 
 \leq \epsilon^{\cP}_D + (1/2)((4 \ln 2) I(X:M))^{1/2}$.

We now give the details of the proof.
Let $\rho^x$ be the density matrix of the first message $M$ of
protocol $\cP$ when Alice's input $X=x$. 
Let $A$ and $B$ 
denote Alice's work qubits excluding the qubits of $M$
and Bob's work qubits respectively. 
Let $\ket{\psi^x}_{AM}$ be the joint pure state of $AM$
in $\cP$ when Alice's input $X=x$, she has finished preparing
her first message, but no communication has taken place as yet.
Without loss of generality, the number of 
qubits in $A$ is at least the number of qubits in $M$.
Define $\rho \defeq \sum_x p_x \rho^x$, where
$p_x$ is the marginal probability of $X=x$ under distribution $D$.
$\rho$ is the density
matrix of the average first message of $\cP$ under distribution $D$. 

\paragraph{Stage 1:} 
We construct a $[t;c;l_1,\ldots,l_t]^A$ safe coinless quantum
protocol $\cP'$ for $f$ with average error under distribution $D$
$\epsilon^{\cP'}_D \leq \epsilon^{\cP}_D + 
 (1/2)((4 \ln 2) I(X:M))^{1/2}$,
and where the density matrix of
the first message is independent of the input $x$ to Alice.
We now describe the protocol $\cP'$.
Suppose Alice is given
$x \in \cX$ and Bob is given $y \in \cY$.
The qubits of $AM$ are initialised to zero. Alice applies
a unitary transformation $U'_x$ on $AM$
in order to prepare a purification $\ket{\phi^x}$
of $\rho$. She then sends the qubits of $M$ as her first message
to Bob. After this, Alice and Bob behave as in protocol $\cP$.
Hence in $\cP'$ the density matrix of the first message
is independent of Alice's input.

Let us now compare the situation in protocols
$\cP$ and $\cP'$ when Alice's input is $x$, Bob's input is $y$,
Alice has finished preparing the state $\ket{\phi^x}$, but no
communication has taken place as yet. 
In protocol $\cP$, the state of $A M B$ at this point in time
is $\ket{\psi^x}_{AM} \ket{\bzero}_B$.
In protocol $\cP'$, the state of $A M B$ at this point in time
is $\ket{\phi^x}_{AM} \ket{\bzero}_B$.
Hence, 
$\epsilon^{\cP'}_{x,y} \leq \epsilon^{\cP}_{x,y} +
 (1/2)\ellone{\ketbra{\psi^x} - \ketbra{\phi^x}}$.
Let $q_{xy}$ denote
the probability that $(X,Y)=(x,y)$ under distribution
$D$. Then, the average error of $\cP'$ under distribution $D$
\begin{eqnarray*}
\epsilon^{\cP'}_D 
  &   =  & \sum_{x,y} q_{xy} \epsilon^{\cP'}_{x,y} \\
  & \leq & \sum_{x,y} q_{xy} 
           \left(\epsilon^{\cP}_{x,y} + \frac{1}{2}
                 \ellone{\ketbra{\psi^x} - \ketbra{\phi^x}}
           \right)  \\
  &   =  & \epsilon^{\cP}_D + \frac{1}{2}
           \sum_x p_x \ellone{\ketbra{\psi^x} - \ketbra{\phi^x}} \\
  & \leq & \epsilon^{\cP}_D + \frac{1}{2} ((4 \ln 2) I(X:M))^{1/2} 
\end{eqnarray*}
The last inequality follows by taking $\ket{\phi^x}$ to be the
purifications promised by Fact~\ref{fact:averageencodingquant}.

\paragraph{Stage 2A:} 
We now construct a $[t+1;c+l_1;0,0,l_2,\ldots,l_t]^B$ safe 
coinless quantum protocol $\cQ'$ for $f$ with
$\epsilon^{\cQ'}_{x,y} = \epsilon^{\cP'}_{x,y}$, for all
$(x,y) \in \cX \times \cY$. 
Suppose
Alice is given $x \in \cX$ and Bob is given $y \in \cY$. 
Let $A_1$ denote
all the qubits of $A$, except the last $l_1+c$ qubits.
Let $A_2$ denote the last $l_1+c$ qubits of $A$. Thus, $A = A_1 A_2$.
In protocol $\cQ'$, Alice initially starts with work qubits $A_1$
only, and Bob initially starts with work qubits $B M A_2$.
The work qubits of Alice and Bob are initialised to zero.
Bob commences protocol $\cQ'$ by constructing
a canonical purification $\ket{\eta}_{M A_2}$ of $\rho$, where
the reduced density matrix of $M$ is $\rho$.
Bob then sends $A_2$ to Alice. 
The density matrix of $A_2$ is independent 
of the inputs $x,y$ (in fact, if $\ket{\eta}_{M A_2}$
is the Schmidt purification then the reduced
density matrix of $A_2$ is also $\rho$). 
After receiving $A_2$, Alice applies a unitary transformation
$V_x$ on $A$ so that the state vector of $A M$ becomes
$\ket{\phi^x}_{AM}$. The existence of such
a $V_x$ follows from Fact~\ref{fact:localtransition}.
The global state of Alice's and Bob's qubits at this point
in protocol $\cQ'$ is the same as the global state of
Alice's and Bob's qubits at the point in protocol $\cP'$ just
after Alice has sent her first message to Bob.
Bob now treats $M$ as if
it were the first message of Alice in $\cP'$, and proceeds to compute
his response $N$ (the qubits of $N$ are a subset of the qubits
of $M B$) of length $l_2$. Bob sends $N$ to Alice and 
after this protocol $\cQ'$ proceeds as protocol $\cP'$. In $\cQ'$
Bob starts the communication, the communication
goes on for $t+1$ rounds, the first message of Bob of length
$l_1+c$ viz. $A_2$ is a safe message, and the first
message of Alice is zero qubits long. 

\paragraph{Stage 2B:} 
We finally construct our desired $[t-1;c+l_1;l_2,\ldots,l_t]^B$ safe 
coinless quantum protocol $\cQ$ for $f$ with
$\epsilon^{\cQ}_{x,y} = \epsilon^{\cQ'}_{x,y}$, for all
$(x,y) \in \cX \times \cY$. 
In protocol $\cQ$ Bob, after doing the same computations as 
in $\cQ'$, first sends as a single
message the $(l_1+c)+l_2$ qubits $A_2 N$.
After receiving $A_2 N$,
Alice applies $V_x$ on $A$ followed by her appropriate
unitary transformation 
on $A N$ viz. the unitary transformation of Alice in $\cQ'$ on
the qubits of $A N$ after she has received the first two 
messages of Bob.
The global state of all the qubits of
Alice and Bob at this point in protocol $\cQ$ 
is the same as the global state of all the qubits of Alice
and Bob at the point in protocol $\cQ'$ just after Alice has finished
generating her second message but before she has sent it to Bob.
This is because
unitary transformations on disjoint sets of qubits commute.
After this, protocol $\cQ$ proceeds as protocol $\cQ'$. 
In protocol $\cQ$ Bob starts the communication, the 
communication
goes on for $t-1$ rounds, and the first message of Bob of length
$(l_1+c)+l_2$ viz. $A_2 N$ contains a safe overhead viz. $A_2$ of 
$l_1+c$ qubits. 

This completes the proof of Lemma~\ref{lem:quantroundreduce}.
\end{proof}

\paragraph{Remark:}
The proof of Lemma~\ref{lem:quantroundreduce} requires the global
state of all the qubits of Alice and Bob to be pure at all times
during the execution of protocol $\cP$, when Alice's and Bob's
inputs are in computational basis states. This is because we use
the machinery of purifications in the proof. The proof also ensures
that the purity property holds for the final protocol $\cQ$.

\section{The round elimination lemma}
\label{sec:roundelim}
We now prove the classical and quantum versions of
our strong round elimination lemma in 
Sections~\ref{subsec:classicalroundelim} and 
\ref{subsec:quantroundelim} respectively. The round elimination
lemma is stated for public coin protocols
only. Since a public coin classical randomised protocol can
be converted to a private coin classical randomised
protocol at the expense of an
additive increase in the communication complexity by at most
logarithm of the total bit size of the inputs~\cite{newman:private},
we also get a similar round elimination lemma for private 
coin classical randomised protocols. A similar statement about
round elimination can
be made for safe coinless quantum protocols; for such protocols
the safe overhead increases by an additional additive term
that is at most
logarithm of the total bit size of the inputs.
But since the statement of the round elimination lemma is cleanest
for public coin protocols, we give it below 
for such protocols only. 

\subsection{Classical}
\label{subsec:classicalroundelim}
\begin{lemma}[Round elimination lemma, classical version] 
\label{lem:classicalroundelim}
Suppose $f:\cX \times \cY \rightarrow \cZ$ is a function.
Suppose the communication game $f^{(n),A}$ has a 
$[t;l_1,\ldots,l_t]^A$ public coin randomised protocol
with error less than $\delta$.
Then there is a $[t-1;l_2,\ldots,l_t]^B$ public coin
randomised protocol for $f$ with error less than 
$\epsilon \defeq \delta + (1/2) (2 l_1 \ln 2 /n)^{1/2}$.
\end{lemma}
\begin{proof} 
Suppose the given protocol for $f^{(n),A}$ has error
$\tilde{\delta} < \delta$. Define 
$\tilde{\epsilon} \defeq \tilde{\delta} + 
 (1/2) (2 l_1 \ln 2 /n)^{1/2}$.
To prove the round elimination lemma
it suffices to give,
by the harder direction of Yao's minimax lemma~\cite{yao:minimax},
for any probability distribution $D$ on $\cX \times \cY$, a 
$[t-1;l_2,\ldots,l_t]^B$ deterministic 
protocol $\cP$ for $f$ with 
$\epsilon^\cP_D \leq \tilde{\epsilon} < \epsilon$. 
To this end, we will first construct a probability distribution 
$D^\ast$ on $\cX^n \times [n] \times \cY$ as follows:
Choose $i \in [n]$ uniformly at random. Choose independently, for 
each $j \in [n]$,
$(x_j,y_j) \in \cX \times \cY$ according to distribution $D$. Set
$y=y_i$ and throw away $y_j, j \neq i$.
By the easier direction of Yao's minimax lemma, we get a 
$[t;l_1,\ldots,l_t]^A$ deterministic protocol $\cP^\ast$
for $f^{(n),A}$ with 
$\epsilon^{\cP^\ast}_{D^\ast} \leq \tilde{\delta} < \delta$. 
In $\cP^\ast$, Alice gets $x_1,\ldots,x_n \in \cX$, Bob gets
$i \in [n]$, $y \in \cY$ and copies of $x_1,\ldots,x_{i-1}$.
We shall construct the desired protocol $\cP$
from the protocol $\cP^\ast$.

In $\cP^\ast$, let Alice's and Bob's inputs be distributed according
to $D^\ast$. Let the input to Alice be denoted by
the random variable $X \defeq X_1 \cdots X_n$, where $X_i$ is the 
random variable corresponding to the $i$th input to Alice. 
Let the random variables $Y, \cI$ correspond to the inputs $y, i$
respectively of Bob. 
Let $M$ denote the random variable corresponding to
the first message of Alice in $\cP^\ast$. 
Define probability distribution $D^\ast_{i;x_1,\ldots,x_{i-1}}$ on
$\cX^n \times [n] \times \cY$ to be the distribution $D^\ast$
conditioned on $\cI=i$ and $X_1,\ldots,X_{i-1}=x_1,\ldots,x_{i-1}$. 
Define probability distribution $D^\ast_{i;y;x_1,\ldots,x_i}$ on
$\cX^n \times [n] \times \cY$ to be the distribution $D^\ast$
conditioned on $\cI=i$, $Y=y$ and $X_1,\ldots,X_i=x_1,\ldots,x_i$. 
Let $\epsilon^{\cP^\ast}_{D^\ast;i;x_1,\ldots,x_{i-1}}$ denote
the average 
error of $\cP^\ast$ under distribution 
$D^\ast_{i;x_1,\ldots,x_{i-1}}$.
Using Fact~\ref{fact:classicalaveraging} and the fact that 
under distribution $D^\ast$ 
$X_1,\ldots,X_n$ are independent random variables, we get that 
\begin{equation}
\label{eq:info1classical}
\begin{array}{lrl}
\E_{i,X}\, [I((X_i:M) \mid  X_1,\ldots,X_{i-1}=x_1,\ldots,x_{i-1})] 
&   =  & \displaystyle \E_i\, [I (X_i:M X_1 \cdots X_{i-1})] \\
&   =  & \frac{I(X:M)}{n} \\
& \leq & \frac{l_1}{n}. 
\end{array}
\end{equation}
Also,
\begin{equation}
\label{eq:error1classical}
\tilde{\delta} \geq \epsilon^{\cP^\ast}_{D^\ast} =
    \E_{i,X}\, \left[\epsilon^{\cP^\ast}_{D^\ast;i;x_1,\ldots,x_{i-1}}
	       \right].
\end{equation}
Above, the expectations are under distribution $D^\ast$ and the
mutual informations are for protocol $\cP^\ast$ with its inputs
distributed according to $D^\ast$.

For any $i \in [n]$, $x_1, \ldots, x_{i-1} \in \cX$, let us now define
the $[t;l_1,\ldots,l_t]^A$
private coin randomised protocol $\cP'_{i;x_1,\ldots,x_{i-1}}$
for the function $f$ in terms of protocol $\cP^\ast$ as follows:
Alice is given
$x \in \cX$ and Bob is given $y \in \cY$. Bob sets
$\cI=i$, and both Alice and Bob set
$X_1, \ldots, X_{i-1} = x_1, \ldots, x_{i-1}$. 
Alice tosses a fresh private coin
to choose $X_{i+1},\ldots,X_n \in \cX$, where each
$X_j, i+1 \leq j \leq n$ is chosen independently according to the
marginal distribution on $\cX$ induced by $D$.
Alice sets $X_i=x$ and Bob sets 
$Y=y$. They then run protocol $\cP^\ast$ on these inputs. 
The probability that $\cP'_{i;x_1,\ldots,x_{i-1}}$
makes an error for an input $(x,y)$,
$\epsilon^{\cP'_{i;x_1,\ldots,x_{i-1}}}_{x,y}$,
is the average probability of
error of $\cP^\ast$ under distribution $D^\ast_{i;y;x_1,\ldots,x_i}$.
Hence, the average probability of error of 
$\cP'_{i;x_1,\ldots,x_{i-1}}$ 
under distribution $D$,
\begin{equation}
\label{eq:error2classical}
\epsilon^{\cP'_{i;x_1,\ldots,x_{i-1}}}_D = 
       \epsilon^{\cP^\ast}_{D^\ast;i;x_1,\ldots,x_{i-1}}.
\end{equation}
Let $M'$ denote the random variable corresponding to Alice's 
first message 
and $X'$ denote the random variable $X_i$, when Alice's and Bob's
inputs are distributed according
to $D$ in $\cP'_{i;x_1,\ldots,x_{i-1}}$.
Then
\begin{equation}
\label{eq:info2classical}
I((X_i:M) \mid  X_1,\ldots,X_{i-1}=x_1,\ldots,x_{i-1}) = I (X':M'),
\end{equation}
where the left hand side refers to
the mutual information in protocol $\cP^\ast$ when its inputs are
distributed according to $D^\ast$ and 
the right hand side refers to the mutual information in
protocol $\cP'_{i;x_1,\ldots,x_{i-1}}$ when its inputs are distributed
according to $D$.

Using Lemma~\ref{lem:classicalroundreduce} and equations 
(\ref{eq:error2classical}) and (\ref{eq:info2classical}), we get a 
$[t-1;l_2,\ldots,l_t]^B$ 
deterministic protocol $\cP_{i;x_1,\ldots,x_{i-1}}$ for 
$f$ with
\begin{equation}
\label{eq:average1classical}
\begin{array}{lcl}
\epsilon^{\cP_{i;x_1,\ldots,x_{i-1}}}_D
& \leq & \epsilon^{\cP'_{i;x_1,\ldots,x_{i-1}}}_D +
         \frac{1}{2} ((2 \ln 2) I(X':M'))^{1/2} \\
&   =  & \epsilon^{\cP^\ast}_{D^\ast;i;x_1,\ldots,x_{i-1}} +
         \frac{1}{2} ((2 \ln 2) 
                      I((X_i:M) \mid X_1,\ldots, X_{i-1} =
                                     x_1,\ldots,x_{i-1}))^{1/2}.
\end{array}
\end{equation}
We have that
(note that the expectations below are under distribution $D^\ast$ and
the mutual informations are for protocol $\cP^\ast$ with its
inputs distributed according to $D^\ast$)
\begin{eqnarray*}
\E_{i,X} \left[\epsilon^{\cP_{i;x_1,\ldots,x_{i-1}}}_D\right]
& \leq & \E_{i,X} \left[\epsilon^{\cP^\ast}_{D^\ast;i;
                                             x_1,\ldots,x_{i-1}}
                  \right] + \\
&      & \frac{1}{2} \E_{i,X}\, \left[
         ((2 \ln 2) I((X_i:M) \mid X_1,\ldots,X_{i-1} = 
                                   x_1,\ldots,x_{i-1}))^{1/2}
         \right] \\
& \leq & \E_{i,X} \left[\epsilon^{\cP^\ast}_{D^\ast;i;
                                             x_1,\ldots,x_{i-1}}
                  \right] + \\
&      & \frac{1}{2} \left(
         (2 \ln 2) \E_{i,X} 
         [I((X_i:M) \mid X_1,\ldots, X_{i-1}=
                         x_1,\ldots,x_{i-1})]
         \right)^{1/2} \\
& \leq & \tilde{\delta} + \frac{1}{2} 
         \left(\frac{2 l_1 \ln 2}{n} \right)^{1/2} \\
&   =  & \tilde{\epsilon}.
\end{eqnarray*}
The first inequality follows from (\ref{eq:average1classical}), 
the second
inequality follows from the concavity of the square root function
and the last inequality from (\ref{eq:info1classical}) and
(\ref{eq:error1classical}).

Thus, we can immediately 
see that there exist $i \in [n]$ and 
$x_1, \ldots, x_{i-1} \in \cX$ such that 
$\epsilon^{\cP_{i;x_1,\ldots,x_{i-1}}}_D \leq \tilde{\epsilon}$. 
Let $\cP \defeq \cP_{i;x_1,\ldots,x_{i-1}}$. $\cP$ is 
our desired $[t-1;l_2,\ldots,l_t]^B$ 
deterministic protocol for $f$ with 
$\epsilon^\cP_D \leq \tilde{\epsilon}$, thus completing the proof
of the round elimination lemma.
\end{proof}

\subsection{Quantum}
\label{subsec:quantroundelim}
\begin{lemma}[Round elimination lemma, quantum version] 
\label{lem:quantroundelim}
Suppose $f:\cX \times \cY \rightarrow \cZ$ is a function.
Suppose the communication game $f^{(n),A}$ has a 
$[t;c;l_1,\ldots,l_t]^A$ safe public coin quantum protocol
with error less than $\delta$.
Then there is a $[t-1;c+l_1;l_2,\ldots,l_t]^B$ safe public coin
quantum protocol for $f$ with error less than 
$\epsilon \defeq \delta + (1/2) (8 l_1 \ln 2 /n)^{1/2}$.
\end{lemma}
\begin{proof} {\bf (Sketch)}
The proof is very similar to the proof
of Lemma~\ref{lem:classicalroundelim}.
We just point out some important things below. Note
that in the quantum setting, the upper bound in 
(\ref{eq:info1classical}) is 
$\frac{2 l_1}{n}$ by Fact~\ref{fact:quantaveraging}(c).
Also, in the definition of protocol $\cP'_{i;x_1,\ldots,x_{i-1}}$,
instead of feeding probabilistic mixtures for the inputs
$X_{i+1}, \ldots, X_n$, Alice feeds appropriate pure states 
(pure states that would give the correct probabilistic mixture
were they to be measured in the computational basis).
Equations~\ref{eq:error2classical} and \ref{eq:info2classical} 
continue to hold because protocol $\cP^\ast$ is secure. 
Since the global
state of all the qubits of Alice and Bob is pure at all times
during the execution of protocol $\cP'_{i;x_1,\ldots,x_{i-1}}$ 
when Alice's and Bob's
inputs are in computational basis states, 
Lemma~\ref{lem:quantroundreduce} can now be used to get the 
$[t-1;c+l_1;l_2,\ldots,l_t]^B$ safe coinless quantum protocol
$\cP_{i;x_1,\ldots,x_{i-1}}$ for $f$.
\end{proof}

\section{Optimal lower bounds for predecessor}
\label{sec:predlb}
In this section, we prove our (optimal) lower bounds on the query 
complexity of static
predecessor searching in the 
cell probe model with randomised or address-only quantum query
schemes.

\begin{theorem}
\label{thm:classicalpredlb}
Suppose there is a $(n^{O(1)}, (\log m)^{O(1)}, t)$ randomised 
cell probe
scheme for the $(m, n)$-static predecessor problem with error 
probability less than $1/3$.
Then, 
\begin{enumerate}
\item[(a)]
$t = \Omega \left( \frac{\log \log m}{\log \log \log m} \right)$ 
as a function of $m$; 
\item[(b)]
$t = \Omega \left( \sqrt{\frac{\log n}{\log \log n}} \, \right)$ 
as a function of $n$.
\end{enumerate}
The same lower bound also holds for address-only quantum cell 
probe schemes for static predecessor searching.
\end{theorem}
\begin{proof}
The proof is similar to the proof of the lower bound for predecessor
in \cite{miltersen:roundelim},
but with different parameters, and using our stronger round
elimination lemma in its classical version
(Lemma~\ref{lem:classicalroundelim}).

By Fact~\ref{fact:pred2rank}, it suffices to consider 
communication protocols for the
rank parity communication game $\PAR_{\log m, n}$.
Let $n = 2^{(\log \log m)^2 / \log \log \log m}$. 
Let $c_1 \defeq (2 \ln 2) 6^2$.
For any given constants $c_2, c_3 \geq 1$, define
\begin{displaymath}
a \defeq c_2 \log n ~~~~~~~~~~ b \defeq (\log m)^{c_3} 
\end{displaymath}
\begin{displaymath}
t \defeq \frac{\log \log m}{(c_1 + c_2 + c_3) \log \log \log m}.
\end{displaymath}
We will show that $\PAR_{\log m, n}$ does not have 
$(2t,a,b)^A$ public coin randomised communication
protocols with error less than $1/3$, 
thus proving both the desired lower bounds for the 
predecessor problem.

Given a $(2t,a,b)^A$ public coin protocol for 
$\PAR_{\log m, n}$ with error probability at most $\delta$,
we can get a $(2t,a,b)^A$ public coin protocol for 
$\PAR^{(c_1 a t^2),A}_{\frac{\log m}{c_1 a t^2}, n}$
with error probability at most $\delta$
by Fact~\ref{fact:rankred1}. Using 
Lemma~\ref{lem:classicalroundelim}, we get a 
$(2t-1,a,b)^B$ public coin protocol for
$\PAR_{\frac{\log m}{c_1 a t^2}, n}$,
but the error probability increases to at most $\delta + (12t)^{-1}$.
By Fact~\ref{fact:rankred2}, we get
a $(2t-1,a,b)^B$ public coin protocol for
$\PAR^{(c_1 b t^2),B}_{\frac{\log m}{c_1 a t^2} - 
                     \log (c_1 b t^2) - 1, \frac{n}{c_1 b t^2}}$
with error probability at most $\delta + (12t)^{-1}$.
From the given values of the parameters, we see that
$\frac{\log m}{(2 c_1 a t^2)^t} \geq \log (c_1 b t^2) + 1$.
This implies that we also have a $(2t-1,a,b)^B$ public 
coin protocol for
$\PAR^{(c_1 b t^2),B}_{\frac{\log m}{2 c_1 a t^2}, 
\frac{n}{c_1 b t^2}}$
with error probability at most $\delta + (12t)^{-1}$.
Using Lemma~\ref{lem:classicalroundelim} again, we get a 
$(2t-2,a,b)^A$ public coin protocol for
$\PAR_{\frac{\log m}{2 c_1 a t^2}, \frac{n}{c_1 b t^2}}$,
but the error probability increases to at most 
$\delta + 2 (12t)^{-1}$.

We do the above steps repeatedly. We start off with a 
$(2t,a,b)^A$ public coin protocol for
$\PAR_{\log m, n}$ with error probability less than $1/3$. 
After applying the above steps
$i$ times, we get a 
$(2t-2i,a,b)^A$ public coin protocol for
$\PAR_{\frac{\log m}{(2 c_1 a t^2)^i}, \frac{n}{(c_1 b t^2)^i}}$
with error probability less than $1/3 + 2i (12t)^{-1}$.

By applying the above steps $t$ times, we finally get a 
$(0,a,b)^A$ public coin randomised protocol for the problem
$\PAR_{\frac{\log m}{(2 c_1 a t^2)^t}, \frac{n}{(c_1 b t^2)^t}}$
with error probability less than $1/3 + 2t (12t)^{-1} = 1/2$.
From the given values of the parameters, we see that
$\frac{\log m}{(2 c_1 a t^2)^t} \geq (\log m)^{\Omega(1)}$ and
$\frac{n}{(c_1 b t^2)^t} \geq n^{\Omega(1)}$.
Thus, we get a zero round protocol for a rank parity problem on a
non-trivial domain with error probability less than $1/2$, which
is a contradiction.

In the above proof, we are tacitly ignoring ``rounding off'' problems.
We remark that this does not affect the correctness of the proof.

Finally we observe that by using Lemma~\ref{lem:quantroundreduce},
one can prove that the same lower bound holds for address-only
quantum cell probe schemes for static predecessor.
\end{proof}

\section{The `greater-than' problem}
\label{sec:gt}
We illustrate another application of the round elimination
lemma to communication complexity by proving
improved rounds versus communication tradeoffs for the 
`greater-than' problem. 
\begin{theorem}
The bounded error public coin randomised $t$-round
communication complexity of $\GT_n$ 
is lower bounded by $\Omega(n^{1/t}t^{-2})$.
For bounded error quantum protocols with input-independent
prior entanglement for
$\GT_n$, we have a lower bound of $\Omega(n^{1/t}t^{-1})$.
\end{theorem}
\begin{proof}
We recall the following reduction from $\GT_{n/k}^{(k),A}$ to 
$\GT_n$ (see \cite{miltersen:roundelim}): In $\GT_{n/k}^{(k),A}$, 
Alice is given 
$x_1,\ldots,x_k \in \{0,1\}^{n/k}$, Bob is given $i \in [k]$,
$y \in \{0,1\}^{n/k}$, and copies of $x_1,\ldots,x_{i-1}$, and
they have to communicate and decide if $x_i > y$. To reduce 
$\GT_{n/k}^{(k),A}$ to $\GT_n$, Alice constructs 
$\hat{x} \in \{0,1\}^n$ by concatenating $x_1,\ldots,x_k$, 
Bob constructs $\hat{y} \in \{0,1\}^n$ by concatenating 
$x_1,\ldots,x_{i-1},y,1^{n(1 - i/k)}$. It is
easy to see that $\hat{x} > \hat{y}$ iff $x_i > y$.

Suppose there is a $t$-round bounded error public
coin randomised protocol for $\GT_n$ with communication 
complexity $c$.
We can think of the protocol as a $(t,c,c)^A$ public coin randomised
protocol with error probability less than $1/3$.
Suppose $n \geq k^t$, where $k \defeq (2 \ln 2) (3 t)^2 c$.
Applying the self-reduction and Lemma~\ref{lem:classicalroundelim}
alternately for $t$ stages gives us a zero round protocol
for the `greater-than' problem on a non-trivial domain 
with error probability less than $1/2$, 
which is a contradiction. 
Thus, $c = \Omega(n^{1/t} t^{-2})$. 

In the above proof, we are tacitly ignoring ``rounding off'' problems.
We remark that this does not affect the correctness of the proof.

Finally, we observe that for bounded error quantum protocols with
input-independent prior entanglement for $\GT_n$, one can
improve the lower bound to $\Omega(n^{1/t} t^{-1})$ by exploiting
the fact that by definition, a quantum protocol sends fixed length
messages independent of the input.
\end{proof}

\section{Conclusion and open problems}
\label{sec:conclusion}
In this paper, we proved a lower bound for the randomised and
address-only quantum query
complexity of a cell probe scheme for the static predecessor
searching problem. Our lower bound matches the deterministic
cell probe upper bound of Beame and Fich. 
We proved our lower bound by proving a strong
round elimination lemma in communication 
complexity. Our round elimination lemma improves on the round
elimination lemma of Miltersen, Nisan, Safra and Wigderson, and is
crucial to proving our optimal lower bound for predecessor searching.
Our strong round elimination lemma also gives us improved 
rounds versus communication 
tradeoffs for the `greater-than' problem. 
We believe that
our round elimination lemma is of independent interest and should have
other applications.
In fact recently, Chakrabarti and Regev~\cite{chakrabarti:ann} 
have proved an optimal lower bound for randomised cell probe schemes 
for the {\em approximate nearest neighbour} searching problem 
on the Hamming cube $\{0, 1\}^d$. Their proof uses the classical
version of our round
elimination lemma and combines it with a {\em message switching}
argument, drawing on the message compression
ideas of Jain, Radhakrishnan and Sen~\cite{jain:compress},
for classical communication protocols that 
further exploits the
asymmetry in the message lengths of Alice of Bob. 

We believe
that our work brings out an interesting fact. Sometimes,
in order to prove lower bound results, it helps to work in a more
general model of computation. Of course, this 
makes the task of proving
lower bounds harder, but also we now 
have more tools and techniques at 
our disposal. This sometimes enables us to attack the problem
in a clearer fashion, without letting irrelevant details about the
restricted model distract us.
In our case, our attempt to prove a lower bound result in the more
general address-only quantum cell probe model led us
to make better use of powerful tools from information theory,
which finally enabled us to prove optimal lower bounds for predecessor
in the randomised cell probe model! Also, the 
information-theoretic approach gives us a simpler
and clearer lower bound proof as compared to previous lower bound
proofs for predecessor.

The lower bound for predecessor searching for quantum cell probe
schemes works only if the query scheme 
is {\em address-only}. If the query scheme is not address-only, the
asymmetry in the message lengths of Alice and Bob in the 
corresponding quantum communication protocol breaks down.
For a general quantum query scheme, it is an open problem
to prove non-trivial lower bounds for static data structure problems.

The {\em message switching} idea of Chakrabarti 
and Regev~\cite{chakrabarti:ann}
works for classical communication protocols only. Thus, their lower
bound for approximate nearest neighbour searching 
on the Hamming cube
holds for classical cell probe schemes only. Our round elimination
lemma alone does not seem to be able to fully exploit the asymmetry
in the message lengths of Alice and Bob in a communication protocol
for this problem. Proving a non-trivial lower bound for this
problem in the address-only quantum cell probe model remains an
open problem.

\section*{Acknowledgements}
We thank Rahul Jain, Hartmut
Klauck and Peter Bro Miltersen for helpful discussions and feedback,
and Amit Chakrabarti for sending us a copy of \cite{chakrabarti:ann}.
We also thank Ashwin Nayak for helpful discussions and for
pointing out the reference \cite{maneva:interaction},
and Jaikumar Radhakrishnan for reading an early draft of 
\cite{sen:quantcell} and for enlightening discussions.

\bibliography{pred}

\appendix

\section{The average encoding theorem}

\subsection{Classical}
In this subsection, we give a self-contained classical proof,
without using quantum information theory, of 
Fact~\ref{fact:averageencodingclassical}.
We first recall the definition of a classical information-theoretic
quantity called {\em relative entropy}, also known as
{\em Kullback-Leibler divergence}.
\begin{definition}[Relative entropy]
Let $P$ and $Q$ be probability distributions on the same 
sample space $\Omega$. The {\em relative entropy} between
$P$ and $Q$ is defined as
\begin{displaymath}
S(P \| Q) \defeq 
\sum_{x \in \Omega} P(x) \log \left(\frac{P(x)}{Q(x)}\right).
\end{displaymath}
\end{definition}

We now require a non-trivial fact from classical information
theory, which upper bounds the total variation distance of a pair
of probability distributions in terms of their relative entropy.
A proof of the fact can be found in 
\cite[Lemma~12.6.1]{cover:infotheory}.
\begin{fact}
\label{thm:divell1}
Let $P$ and $Q$ be probability distributions on the same finite
sample space $\Omega$. Then,
\[
\ellone{P - Q} \leq \sqrt{(2 \ln 2) S(P \| Q)}.
\]
\end{fact}

We can now prove
Fact~\ref{fact:averageencodingclassical}.

\ \\
{\bf Fact~\ref{fact:averageencodingclassical} (Average encoding
theorem, classical version)}
{\it
Let $X, M$ be correlated 
random variables. Let $p_x$ denote the (marginal) probability
that $X=x$, and $\Pi^x$ denote the conditional distribution of $M$
given that $X=x$.
Let $\Pi$ denote the (marginal) probability distribution of $M$ i.e.
$\Pi = \sum_x p_x \Pi^x$. 
Then,
\begin{displaymath}
\sum_x p_x \ellone{\Pi^x - \Pi} \leq \sqrt{(2 \ln 2) I(X:M)}.
\end{displaymath}
\/} 
\begin{proof} 
Let $\cX$, $\cM$ be the finite ranges of random 
variables $X$, $M$ respectively.
We define two probability distributions $P$, $Q$ on 
$\cX \times \cM$.
In distribution $P$, the probability of 
$(x,m) \in \cX  \times \cM$ is
$p_x \cdot \pi_m^x$, where $\pi_m^x$ is the conditional probability
that $M = m$ given that $X = x$. In distribution $Q$, the
probability of $(x,m) \in \cX \times \cM$ is $p_x \cdot \pi_m$, where
$\pi_m$ is the (marginal) probability that $M=m$
i.e. $\pi_m = \sum_x p_x \pi_m^x$.

It is easy to check that 
$S(P \| Q) = I(X:M)$ and
$\ellone{P - Q} = \sum_x p_x \ellone{\Pi_x - \Pi}$.
The result now follows by applying Fact~\ref{thm:divell1} to $P$
and $Q$.
\end{proof}

\subsection{Quantum}
In this subsection, we give a proof of 
Fact~\ref{fact:averageencodingquant}.
We first recall some basic definitions and facts from quantum 
information theory.
Let $\rho$ and $\sigma$ be density matrices in the
same finite dimensional Hilbert space $\cH$. 
The {\em fidelity} (also called Uhlmann's transition probability
or the Bhattacharya coefficient) of $\rho$ and $\sigma$ is defined as
$B(\rho, \sigma) \defeq \ellone{\sqrt{\rho} \sqrt{\sigma}}$.
The {\em von Neumann relative entropy} between
$\rho$ and $\sigma$ is defined as
$S(\rho \| \sigma) \defeq \Tr (\rho (\log \rho - \log \sigma))$.

Jozsa~\cite{jozsa:fidelity} gave an elementary proof for finite
dimensional Hilbert spaces of the following basic and remarkable
property about fidelity.
\begin{fact}
\label{fact:jozsa}
Let $\rho, \sigma$ be density matrices in the same 
Hilbert space $\cH$. Let $\cK$ be a disjoint
Hilbert space such that $\dim (\cK) \geq \dim (\cH)$. Then
for any purification $\ket{\psi}$ of $\rho$ in $\cH \otimes \cK$,
there exists a purification $\ket{\phi}$ of $\sigma$ 
in $\cH \otimes \cK$ such that
$B(\rho, \sigma) = |\braket{\psi}{\phi}|$.
\end{fact}

For two probability distributions $P, Q$ on the same sample space
$\Omega$, their fidelity is defined as
$B(P, Q) \defeq \sum_{x \in \Omega} \sqrt{P(x) Q(x)}$.
We will need the following result about 
fidelity proved by Fuchs and Caves~\cite{fuchs:fidelity}.
\begin{fact}
\label{fact:fuchscaves}
Let $\rho, \sigma$ be density matrices in the same 
Hilbert space $\cH$. Then
$\displaystyle B(\rho, \sigma)=\inf_{\cM} B(\cM \rho, \cM \sigma)$,
where $\cM$ ranges over POVM's on $\cH$.
In fact, the infimum above can be attained by a complete 
von Neumann measurement on $\cH$.
\end{fact}

The following fundamental fact (see e.g~\cite{nielsen:quant}) 
states that the relative entropy can only decrease on performing
a measurement.
\begin{fact}[Monotonicity of relative entropy]
\label{fact:monorelentropy}
Let $\rho, \sigma$ be density matrices in the same finite
dimensional Hilbert space $\cH$. Let $\cM$ be a
POVM on $\cH$. Then, 
$S(\cM \rho \| \cM \sigma) \leq S(\rho \| \sigma)$.
\end{fact}

We require the following explicit expression for the 
trace distance of two pure states (see e.g.~\cite{nielsen:quant}).
\begin{fact}
\label{fact:trdistpure}
For pure states $\ket{\psi}$ and $\ket{\phi}$ in the same Hilbert
space, 
$\ellone{\ketbra{\psi} - \ketbra{\phi}} = 
 2 \sqrt{1 - |\braket{\psi}{\phi}|^2}$.
\end{fact}

The following information-theoretic fact follows easily from the
definitions.
\begin{fact}
\label{fact:inforelentropy}
Let $X, M$ be quantum systems with some joint density matrix, where
$X$ is a classical random variable. Let
$p_x$ denote the (marginal) probability that $X=x$ 
and $\rho^x$ denote the conditional density matrix
of $M$ given that $X=x$. 
Let $\rho$ denote the reduced density matrix of $M$. Note that
$\rho = \sum_x p_x \rho^x$. 
Then, $I(X:M) = \sum_x p_x S(\rho^x \| \rho)$.
\end{fact}

We now recall the following direct connection between relative
entropy and fidelity observed in the classical setting by
\cite{dacunha:BgeqS}, and in the quantum setting by Klauck (private
communication).
\begin{fact}
\label{fact:BleqS}
Let $\rho$ and $\sigma$ be two density matrices in the same
Hilbert space. Then, 
\begin{displaymath}
1 - B(\rho , \sigma) \leq \frac{(\ln 2) S(\rho \| \sigma)}{2}.
\end{displaymath}
\end{fact}
\begin{proof}
Let $\cM$ be the complete von Neumann measurement that 
achieves the infimum 
in Fact~\ref{fact:fuchscaves}. Let $\Omega$ denote the set of
possible (classical) outcomes of $\cM$.
Define probability distributions $P \defeq \cM \rho$ and
$Q \defeq \cM \sigma$.
From Fact~\ref{fact:monorelentropy} and concavity of the $\log$ 
function it follows that
\begin{displaymath}
\begin{array}{c}
\displaystyle
-\frac{S(\rho \| \sigma)}{2} \leq -\frac{S(P \| Q)}{2} 
= \sum_{x \in \Omega} P(x) \log  \sqrt{\frac{Q(x)}{P(x)}} \\
\displaystyle
\leq \log \sum_{x \in \Omega} \sqrt{Q(x) P(x)} 
= \log B(P, Q) = \log B(\rho,\sigma).
\end{array}
\end{displaymath}
Thus, 
$B(\rho , \sigma) 
\geq 2^{- S(\rho \| \sigma) / 2} 
  =  \exp (- (\ln 2) S(\rho \| \sigma) / 2) 
\geq  1 - ((\ln 2) S(\rho \| \sigma) / 2)$.
\end{proof}

We can now prove
Fact~\ref{fact:averageencodingquant}.

\ \\
{\bf Fact~\ref{fact:averageencodingquant} (Average encoding
theorem, quantum version)}
{\it
Let $X, M$ be quantum systems with some joint density matrix.
Let $X$ be a classical random variable
and $p_x$ denote the (marginal) probability
that $X=x$. Let $\cH$ denote the Hilbert space of $M$
and $\rho^x$ denote the conditional density matrix
of $M$ given that $X=x$. 
Let $\rho$ denote the reduced density matrix of $M$. Note that
$\rho = \sum_x p_x \rho^x$. 
Let $\cK$ denote a disjoint 
Hilbert space such that $\dim(\cK) \geq \dim(\cH)$.
Let $\ket{\psi^x}$ be purifications of $\rho^x$ in
$\cH \otimes \cK$. Then there exist purifications $\ket{\phi^x}$
of $\rho$ in $\cH \otimes \cK$ such that 
\begin{displaymath}
\sum_x p_x \ellone{\ketbra{\psi^x} - \ketbra{\phi^x}} \leq 
\sqrt{(4 \ln 2) I(X:M)}.
\end{displaymath}
\/} \\
{\bf Proof:}
Using the concavity of the square root function,
Facts~\ref{fact:inforelentropy}, 
\ref{fact:jozsa}, \ref{fact:trdistpure}, \ref{fact:BleqS}, 
and the fact that fidelity is always at most $1$, we get
\begin{displaymath}
\begin{array}{c}
\displaystyle
\sqrt{I(X:M)} 
    =    \sqrt{\sum_x p_x S(\rho^x \| \rho)} 
  \geq   \sum_x p_x \sqrt{S(\rho^x \| \rho)} 
  \geq   \sum_x p_x \sqrt{\frac{2 (1 - B(\rho^x, \rho))}{\ln 2}} \\
\displaystyle
  \geq   \sum_x p_x \sqrt{\frac{1 - (B(\rho^x, \rho))^2}{\ln 2}} 
    =    \sum_x p_x \sqrt{\frac{1 - |\braket{\psi^x}{\phi^x}|^2}
                               {\ln 2}} 
    =    \sum_x p_x \frac{\ellone{\ketbra{\psi^x} - \ketbra{\phi^x}}}
                         {\sqrt{4 \ln 2}}.
\end{array}
\end{displaymath}
\qed

\end{document}